\documentclass[fleqn,usenatbib,useAMS]{mnras}

\usepackage{graphicx}	% Including figure files
\usepackage{amsmath}	% Advanced maths commands
\usepackage{amssymb}	% Extra maths symbols
\usepackage{multicol}        % Multi-column entries in tables
\usepackage{bm}		% Bold maths symbols, including upright Greek
\usepackage{pdflscape}	% Landscape pages
\usepackage{csvsimple}

\usepackage[T1]{fontenc}
\usepackage{ae,aecompl}

\usepackage{newtxtext,newtxmath}
\title[RVM fits to MeerKAT pulsar data]
{The Thousand-Pulsar-Array programme on MeerKAT XI: Application of the rotating vector model}
\author[Johnston et al.]
{S.~Johnston$^{1}$\thanks{Email: simon.johnston@csiro.au},
M.~Kramer$^{2}$,
A.~Karastergiou$^{3,4}$,
M.~J.~Keith$^5$,
L.~S.~Oswald$^{3,6}$,\newauthor
A.~Parthasarathy$^{2}$,
P.~Weltevrede$^5$
\\
$^1$Australia Telescope National Facility, CSIRO Space and Astronomy, PO~Box~76, Epping NSW~1710, Australia\\
$^2$Max-Planck-Institut f{\"u}r Radioastronomie, Auf dem H{\"u}gel 69, D-53121 Bonn, Germany\\
$^3$Department of Astrophysics, University of Oxford, Denys Wilkinson Building, Keble Road, Oxford OX1 3RH, UK\\
$^4$Department of Physics and Electronics, Rhodes University, PO Box 94, Grahamstown 6140, South Africa\\
$^5$Jodrell Bank Centre for Astrophysics, Department of Physics and Astronomy, University of Manchester, Manchester M13 9PL, UK\\
$^6$Magdalen College, University of Oxford, Oxford OX1 4AU, UK\\
}
\date{Last updated; in original form}

\pubyear{2022}

\begin{document}
\label{firstpage}
\pagerange{\pageref{firstpage}--\pageref{lastpage}}
\maketitle

\begin{abstract}
In spite of the rich phenomenology of the polarization properties of radio pulsars, the rotating vector model (RVM) created 50 years ago remains the best method to determine the beam geometry of a pulsar. We apply the RVM to a sample of 854 radio pulsars observed with the MeerKAT telescope in order to draw conclusions about the population of pulsars as a whole. The main results are that (i) the geometrical interpretation of the position angle traverse is valid in the majority of the population, (ii) the pulsars for which the RVM fails tend to have a high fraction of circular polarization compared to linear polarization, (iii) emission heights obtained through both geometrical and relativistic methods show that the majority of pulsars must have emission heights less than 1000~km independent of spin period, (iv) orthogonal mode jumps are seen in the position angle traverse in about one third of the population. All these results are weakly dependent on the pulsar spin-down energy.
%Finally, we seem to view 60\% of pulsars on inner (pole-ward) lines of sight, compared to 40\% on outer (equator-ward) lines. This result warrants further investigation.
\end{abstract}


%%%%%%%%%%%%%%%%%%%%%%%%%%%%%%%%%%%%%%%%%%%%%%%%%%

%%%%%%%%%%%%%%%%% BODY OF PAPER %%%%%%%%%%%%%%%%%%
\section{Introduction}
Arguably the most important characteristics of a pulsar are its spin period, $P$, its spin-down energy loss-rate, $\dot{E}$, and its geometry. The offset between the magnetic and rotation axes (denoted by $\alpha$) and the traverse of the observer's line-of-sight determines whether or not a pulsar is visible and the shape and polarization of its pulse profile. Furthermore, the birth distribution of $\alpha$ and its evolution with time are critical to understanding the population of pulsars as a whole and the formation process of neutron stars. $P$ and $\dot{E}$ are trivial to measure, and values exist for some 3000 pulsars. The geometrical angles are much harder to determine and unambiguous values exist for at best a few tens of pulsars.

In the rotating vector model (RVM) of \citet{rc69}, the radio radiation is beamed along the field lines and the plane of polarization is determined by the orientation of the magnetic field as it sweeps past the line of sight. The position angle (PA) as a function of pulse longitude, $\phi$, can be expressed as
\begin{equation}
\label{paswing}
{\rm PA} = {\rm PA}_{0} +
{\rm arctan} \left( \frac{{\rm sin}\alpha
\, {\rm sin}(\phi - \phi_0)}{{\rm sin}\zeta
\, {\rm cos}\alpha - {\rm cos}\zeta
\, {\rm sin}\alpha \, {\rm cos}(\phi - \phi_0)} \right)
\end{equation}
Here, $\zeta=\alpha+\beta$ with $\beta$ being the angle of closest approach of the line of sight to the magnetic axis. $\phi_0$ is the pulse longitude at which the PA is PA$_{0}$, which also corresponds to the PA of the rotation axis projected onto the plane of the sky. 

The textbook picture of radio pulsars has emission arising from close to the surface, at a height $h_{\rm em}$, and is bounded by the open field lines. The half-opening angle of the emission cone, $\rho$, is then given by
\begin{equation}
\label{height}
\rho = \sqrt{\frac{9\,\, \pi \,\, h_{\rm em}}{2\,\,P\,\,c}}
\end{equation}
\citep{ran90} with $P$ the spin period and $c$ the speed of light. In turn, the observed pulse width is related to $\rho$ via the geometry and can be expressed \citep{ggr84} as
\begin{equation}
\label{rho}
{\rm cos}\rho = {\rm cos}\alpha\,\, {\rm cos}\zeta\,\, +\,\, {\rm sin}\alpha\,\, {\rm sin}\zeta\,\, {\rm cos}(W_{10}/2)
\end{equation}
\begin{table*}
\caption{Results of the RVM fitting by class. Column 2 lists the number in each class. Columns 3 and 4 indicate the sign of $\beta$. In columns 5 and 6 we denote as V$_{\rm h}$ those pulsars for which $|V|>L$ in 5 bins or more across the profile and V$_{\rm l}$ those pulsars for which this is not the case. The final three columns indicate the offset between the location of $\phi_0$ and the profile midpoint. See text for details.}
\label{tabtype}
\begin{tabular}{lrrrrrrrr}
\hline
Class & Number  & $\beta<0$ & $\beta>0$ & V$_{\rm h}$ & V$_{\rm l}$ & $\Delta\phi<-1$\degr & $|\Delta\phi|<1$\degr & $\Delta\phi>1$\degr\\
\hline & \vspace{-3mm} \\
RVM & 431 & 224 & 207 & 72 & 359 & 79 & 86 & 266\\
flat & 71 & 43 & 28 & 6 & 65\\
non RVM & 352 & & & 210 & 142 \\
Total &  854 \\
\hline
\end{tabular}
\end{table*}
where $W_{10}$ is the pulse width measured at 10\% of the peak flux (e.g. \citealt{pjk+21}).
Furthermore, relativistic effects are important due to the rapid rotation of the magnetosphere as initially discussed in \citet{bcw91}. They showed that the location of the inflection point of the PA swing is delayed with respect to the centre of the pulse profile by an amount given by
\begin{equation}
\label{eqn:dp}
\Delta_\phi = \frac{8\,\,\, \pi\,\,\, h_{\rm em}}{P\,\,\, c}
\end{equation}
For a pulsar with $P=0.25$~s and $h_{\rm em} = 300$~km, $\rho \approx 14$\degr and for $\alpha$ not too low then $W_{10} \approx 28$\degr\ and $\Delta_\phi \approx 6$\degr. Alternatively, measurements of $\Delta\phi$ and/or $W_{10}$ can help constrain $h_{\rm em}$. This approach, and its pitfalls, are outlined in e.g. \citet{wj08}. It should be noted that the standard picture of a hollow cone of emission \citep{ran83,ran90,lm88,mr02} has come under close scrutiny, and an alternative model which invokes `flux tubes' or `fan-beams' is equally good at reproducing the observed phenomenology \citep{wpz+14,dr15,dyks17,okj19}. From a theoretical perspective, Equation~\ref{height} assumes a circular polar cap and no distortion of the magnetosphere, both of which are questionable (e.g. \citealt{gan04,ym14,lgop19}).

In spite of the simplicity of the geometrical argument for the PA traverse, many open questions remain. In about half of all pulsars, the PA traverse does not conform to the RVM and reasons for this must be sought. The main explanation comes from propagation through the magnetosphere, including refraction of the modes \citep{wsve03,fl04,bp12}, coherent mode mixing \citep{dyks19} and generalised Faraday rotation \citep{km98,ijw19}. In addition, it is likely that the various components of the pulse profile arise at different emission heights, with outer components originating from higher in the magnetosphere \citep{ym14,jk19b,dkl+19} where sweepback of the magnetic field lines might be important \citep{cr12}. Finally, the dipolar field may be offset from the star's centre and/or quadropole or higher-order fields may be important (e.g. \citealt{pet20}). Furthermore, even for pulsars which show PA traverses compatible with the RVM, the narrow duty cycle of the pulse profile makes it difficult to determine a unique solution for $\alpha$ and $\beta$ in the absence of further constraints (e.g. \citealt{rwj15b}).

Although these objections may appear daunting, there is good reason to believe the validity of the RVM in many cases. In PSR~J1906+0746 where the geometry changes with time due to precession of the beam, the RVM tracks the geometrical changes beautifully \citep{dkl+19}. Furthermore, in pulsars with interpulses, RVM fitting reflects the geometry of both the main and interpulse components \citep{ww09,jk19b}. In this paper we therefore attempt RVM fits for a sample of more than 1200 pulsars observed using the MeerKAT telescope. Section~2 briefly outlines the observations and the calibration procedure. Section~3 describes how the RVM fitting was performed and Section~4 details the results of the fitting. In Section 5 we present an analysis of the results and their implications for the population of pulsars as a whole.

\begin{figure*}
\begin{center}
\begin{tabular}{ll}
\includegraphics[width=9.0cm,angle=0]{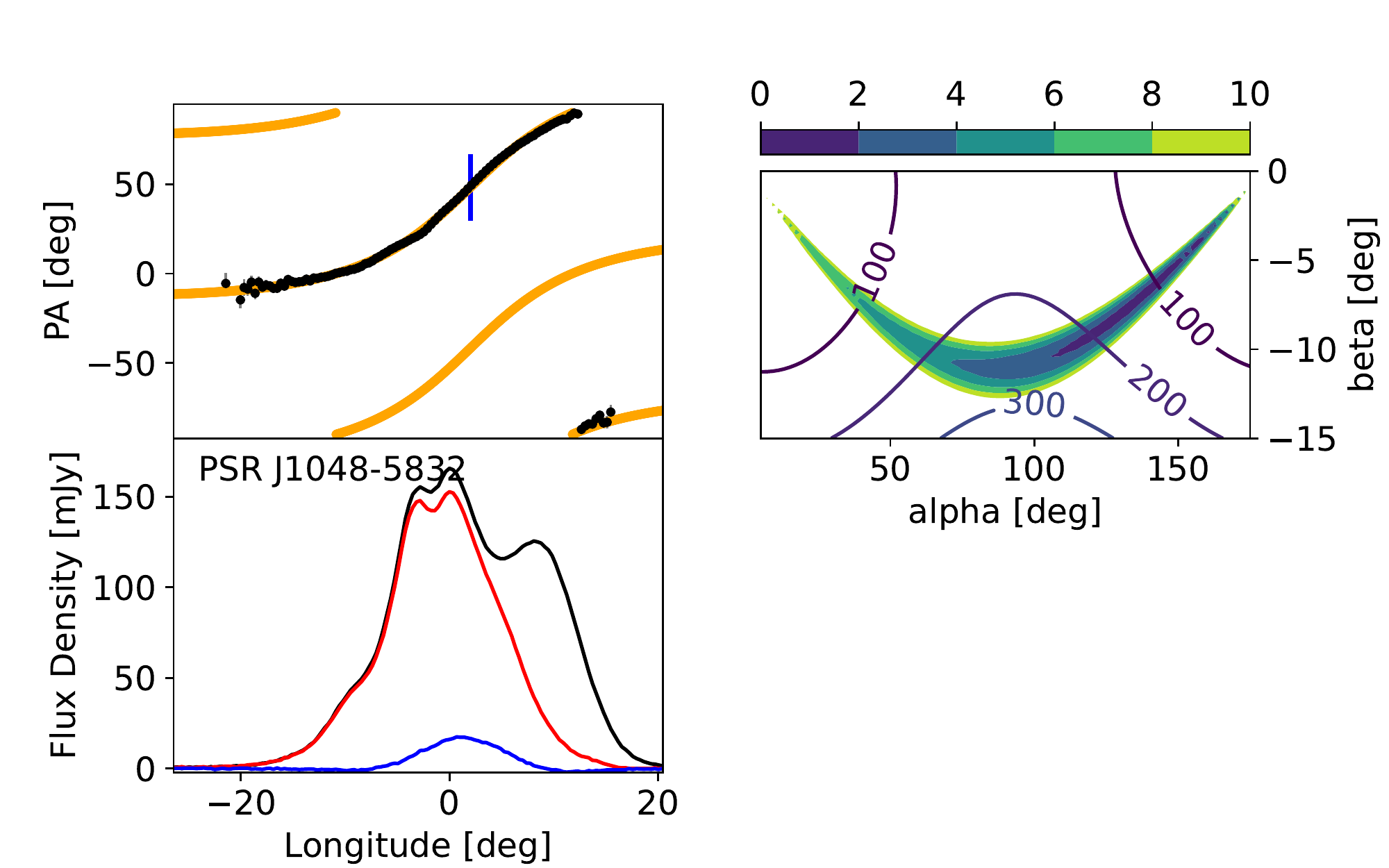} &
\includegraphics[width=9.0cm,angle=0]{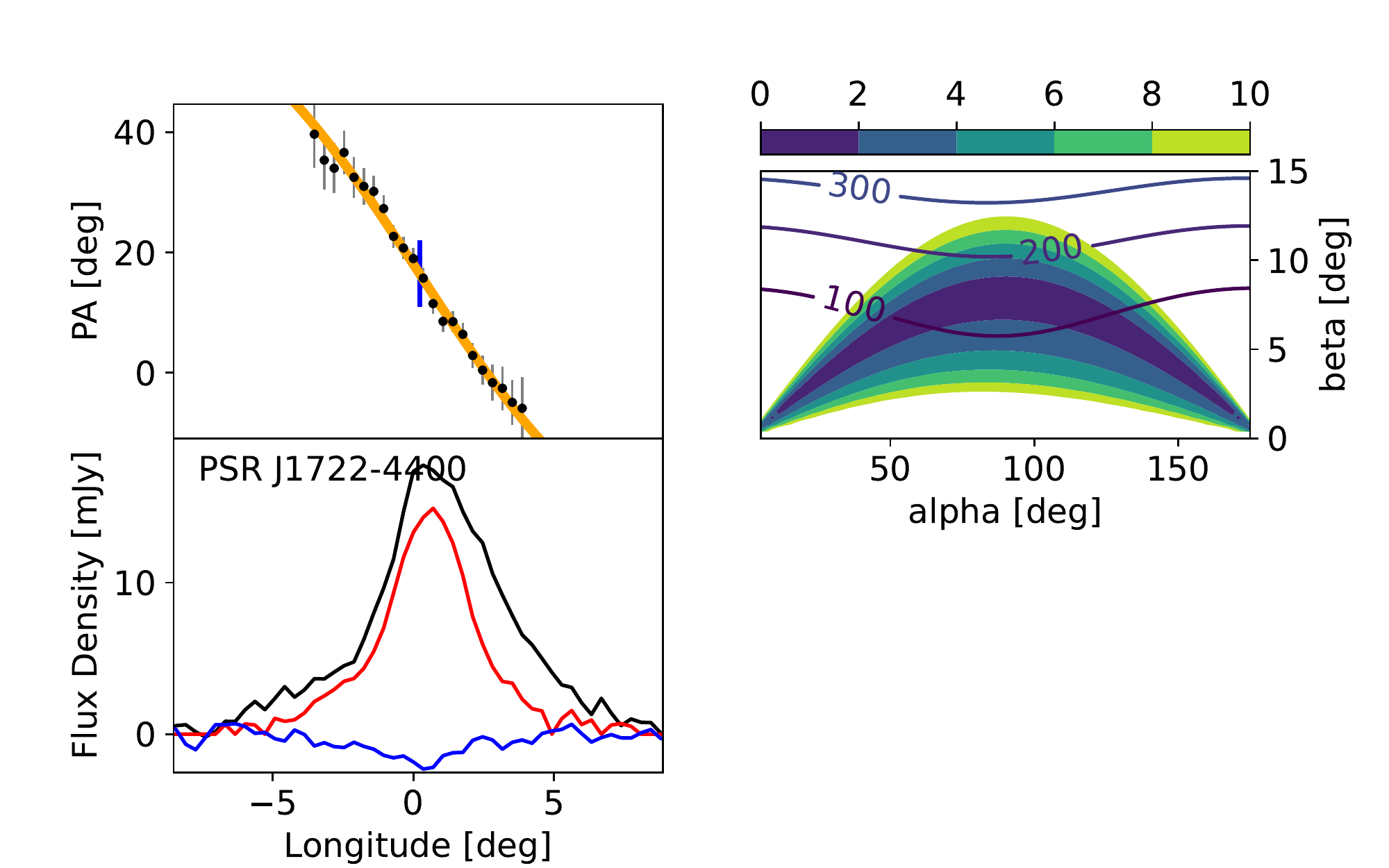} \\
\includegraphics[width=9.0cm,angle=0]{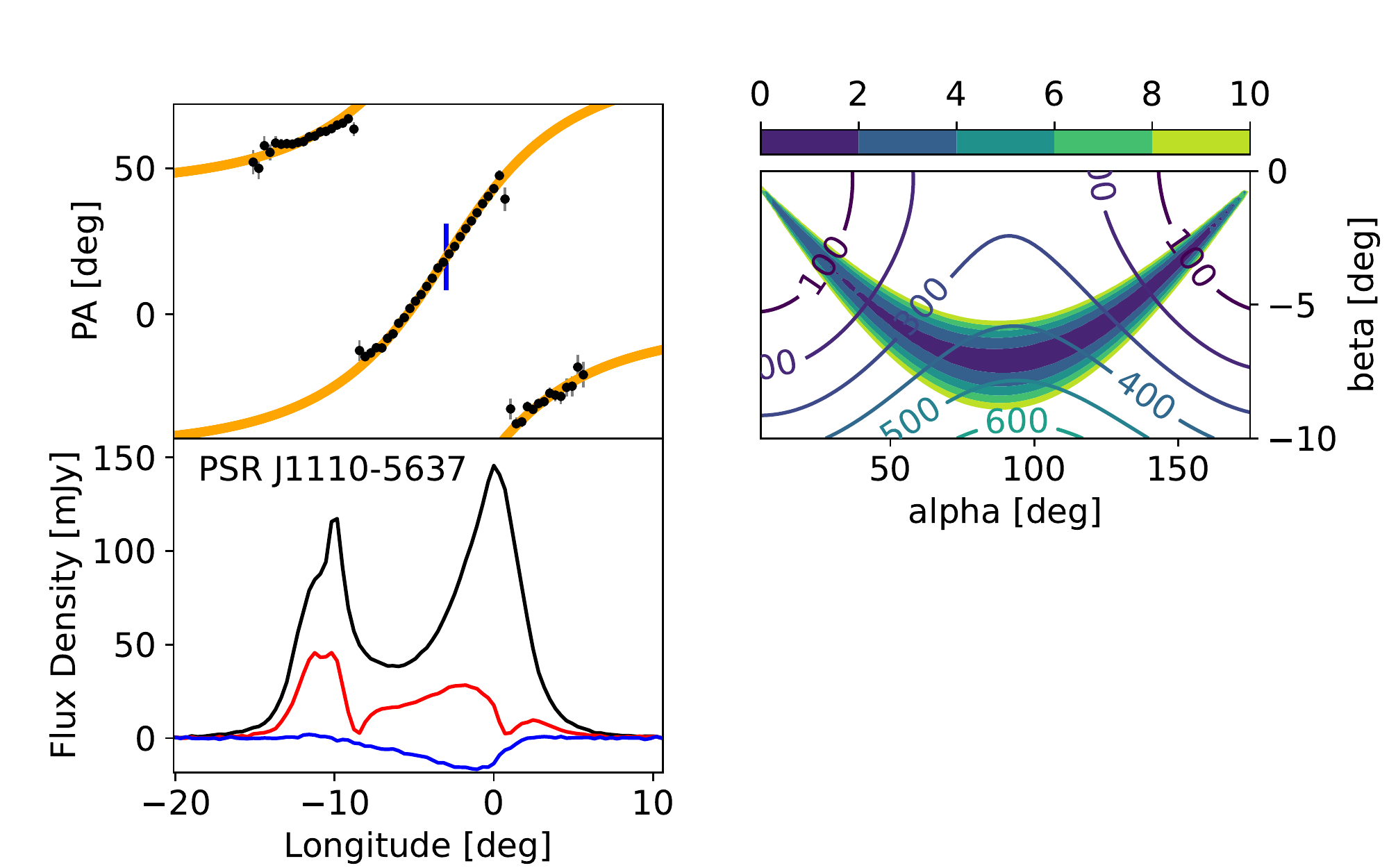} &
\includegraphics[width=9.0cm,angle=0]{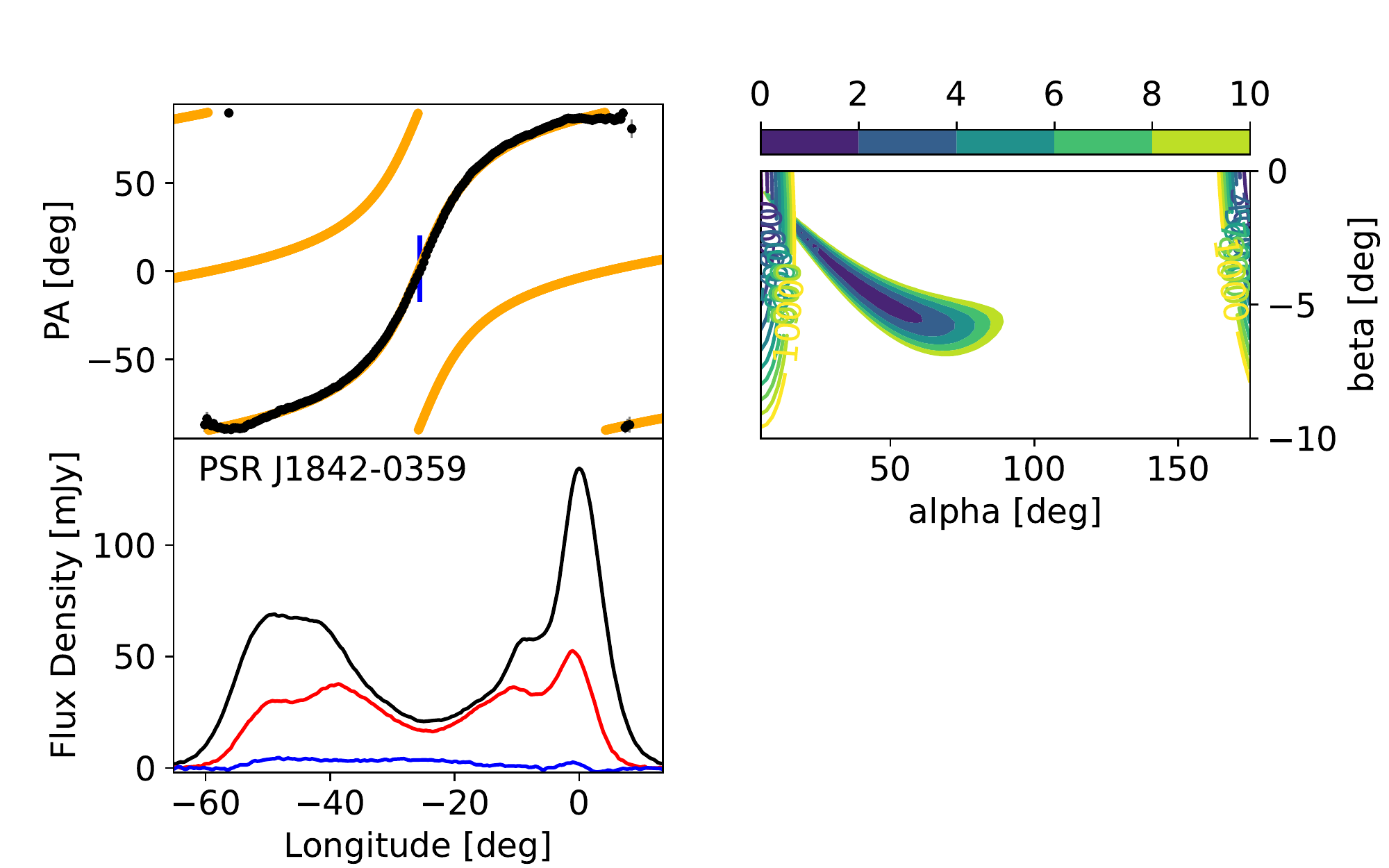} \\
\end{tabular}
\end{center}
\caption{Examples of pulsars in the `RVM' class for which the RVM fitting returns a low minimum value of $\chi^2$. The left-hand panels show the pulsar profile in Stokes I (black), linear polarization (red) and Stokes V (blue). The position angle of the linear polarization is shown in black. The best fit RVM is in orange, along with its 90\degr\ offset. The centre of the blue vertical line marks the inflection point. The right-hand panel shows the $\alpha-\beta$ plane, the image shows values of $\chi^2$ from the fitting routine. Superposed are contours of height in km derived under the assumption that the beam is filled. The pulsars shown are PSRs~J1048--5832, J1722--4400, J1110--5637 and J1842--0359.}
\label{figgood}
\end{figure*}

\begin{figure*}
\begin{center}
\begin{tabular}{ll}
\includegraphics[width=9.5cm,angle=0]{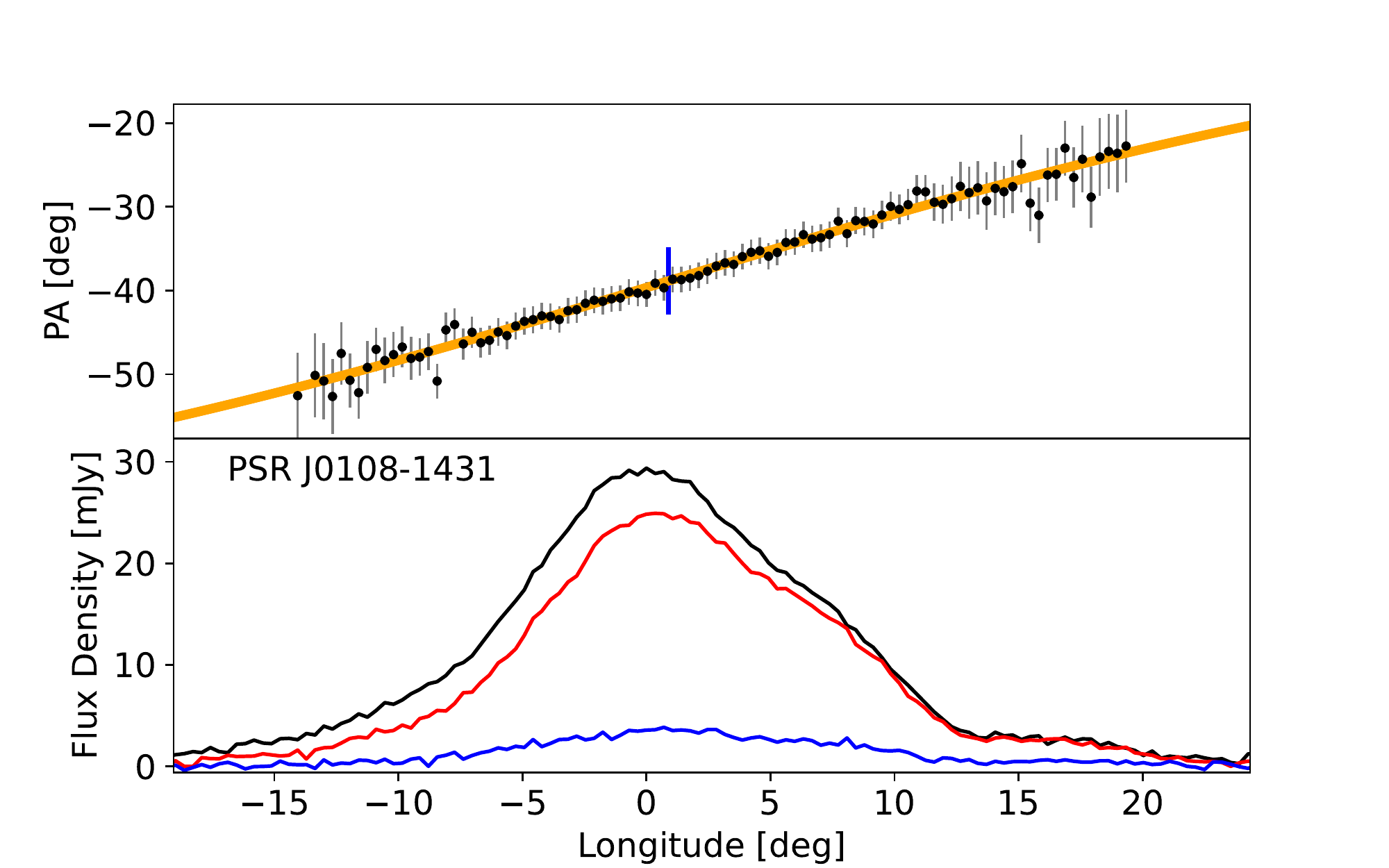} &
\includegraphics[width=9.5cm,angle=0]{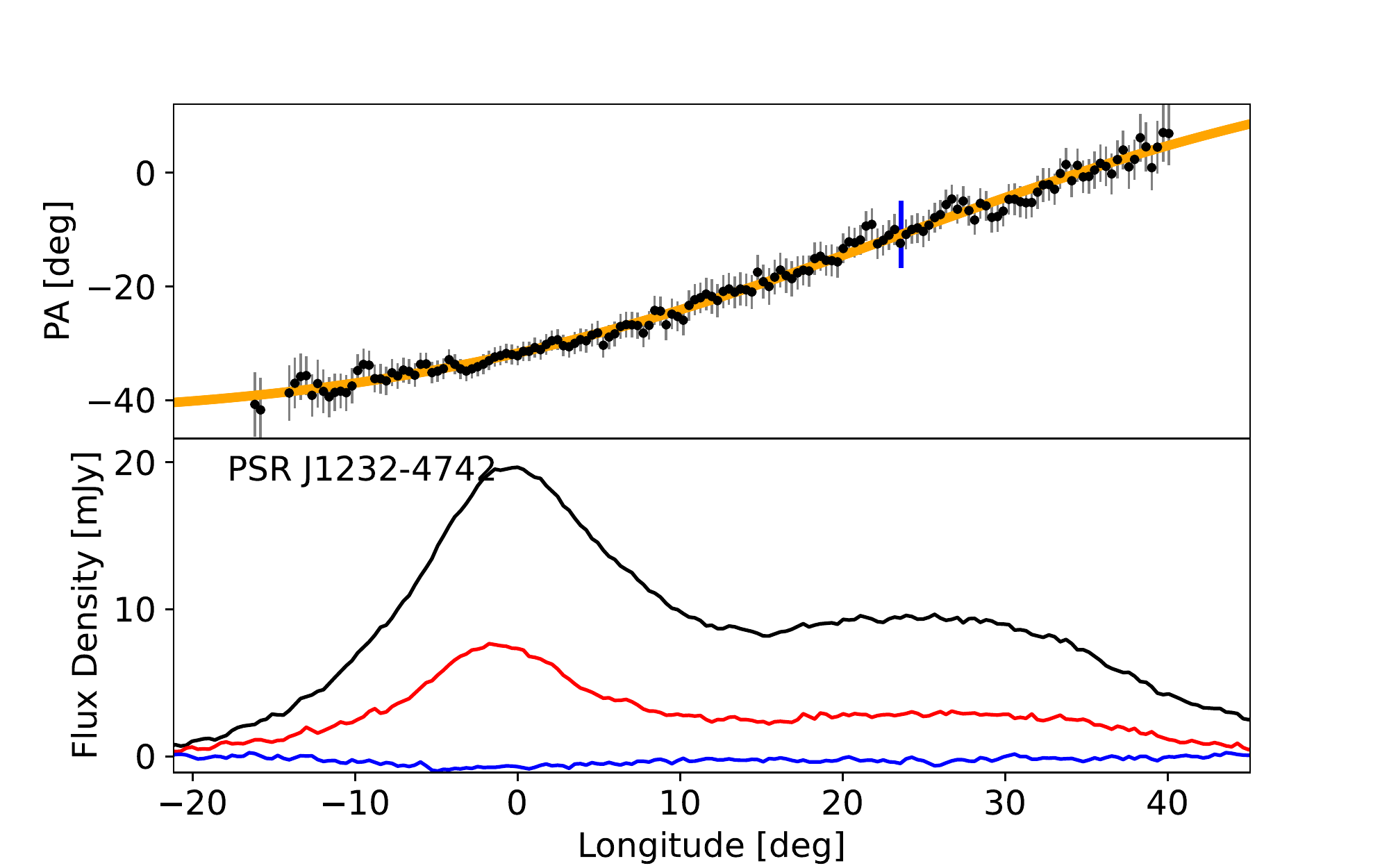} \\
\includegraphics[width=9.5cm,angle=0]{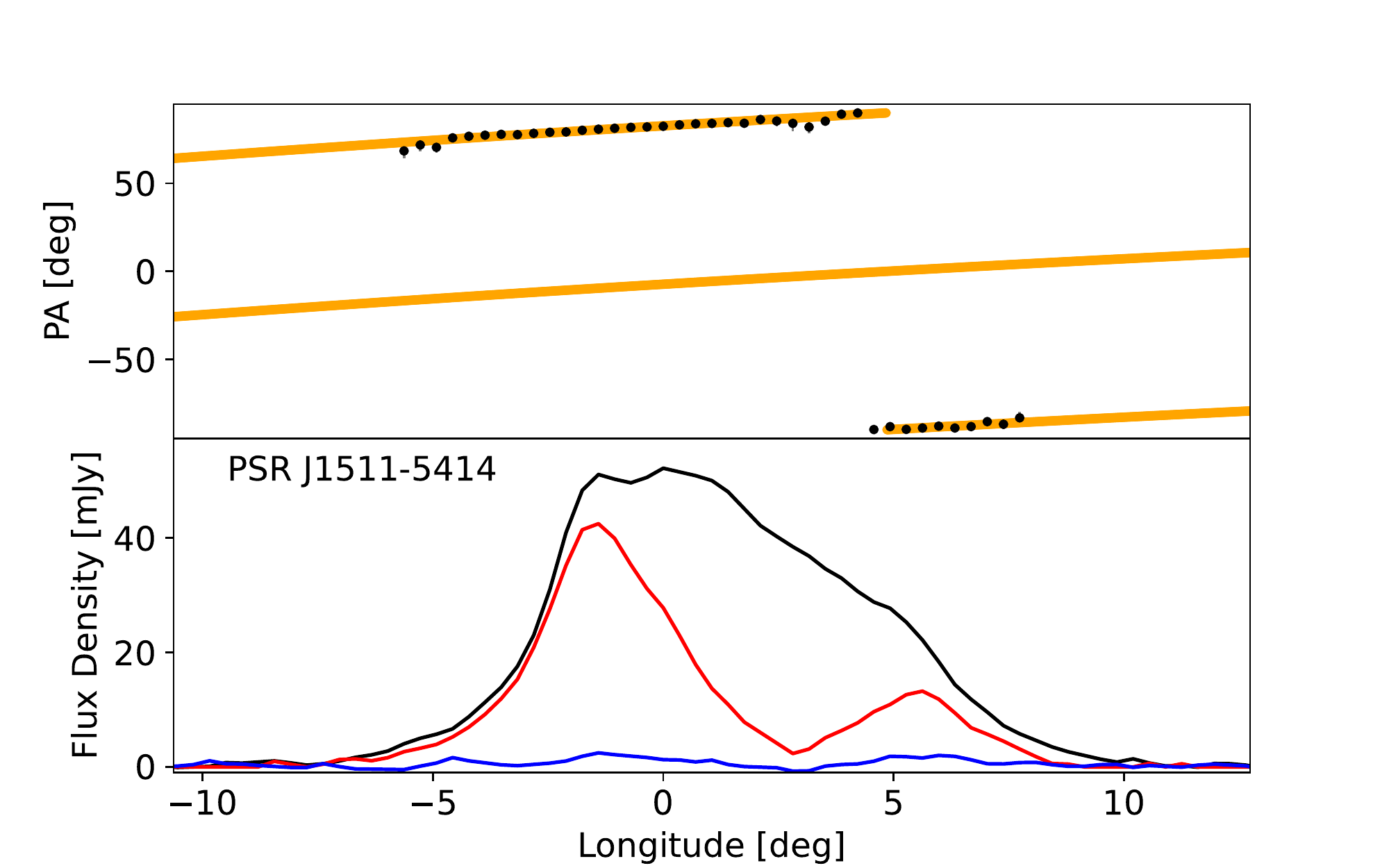} &
\includegraphics[width=9.5cm,angle=0]{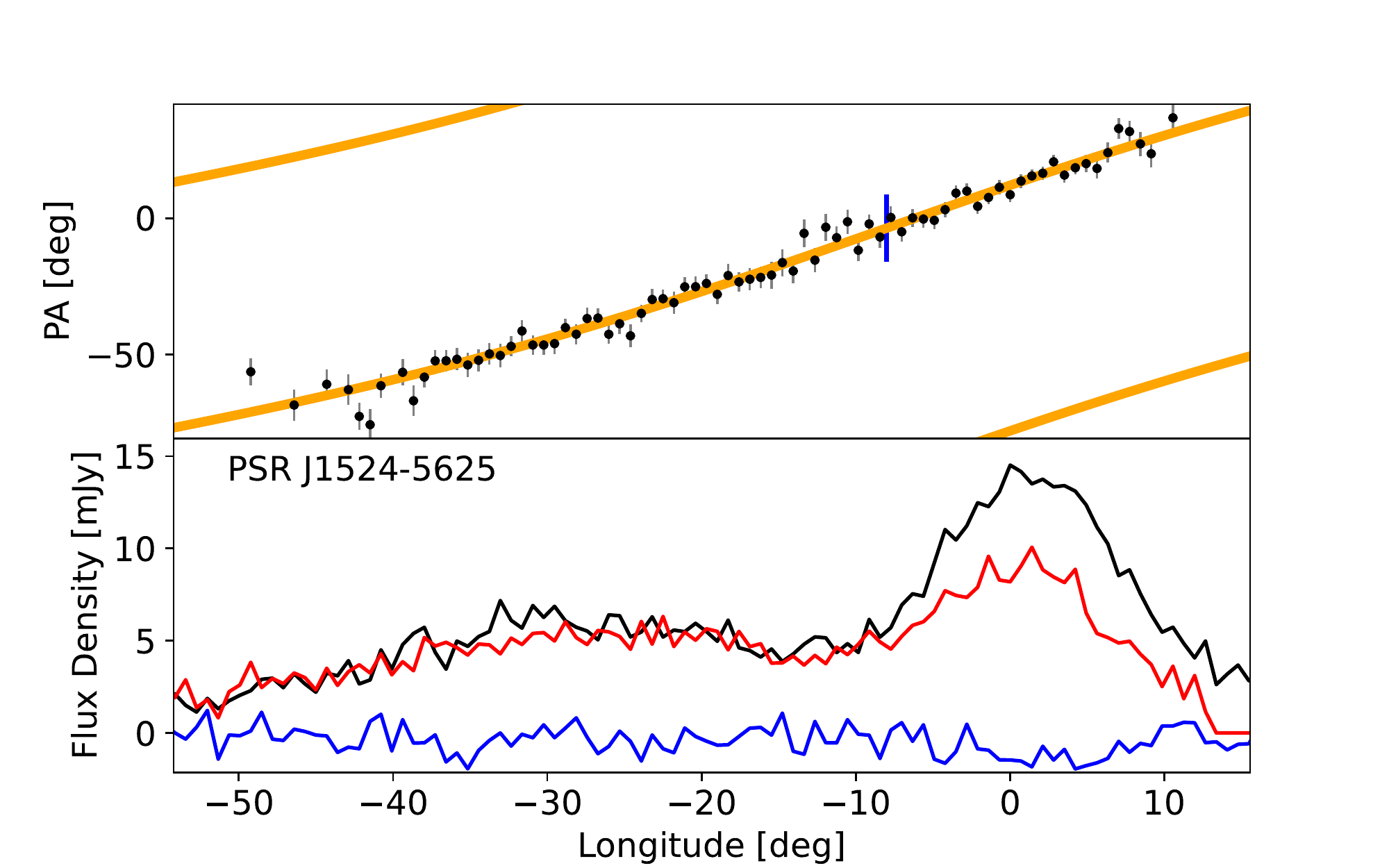} \\
\end{tabular}
\end{center}
\caption{Examples of pulsars in the `flat' class. PSRs~J0108--1431, J1232--4742 and J1511--5414 are pulsars with low values of $\dot{E}$ whereas PSR~J1524--5625 has a high value of $\dot{E}$. All pulsars shown here have a PA slope which is less than 2~deg/deg.}
\label{figflat}
\end{figure*}

\begin{figure*}
\begin{center}
\begin{tabular}{ll}
\includegraphics[width=9.5cm,angle=0]{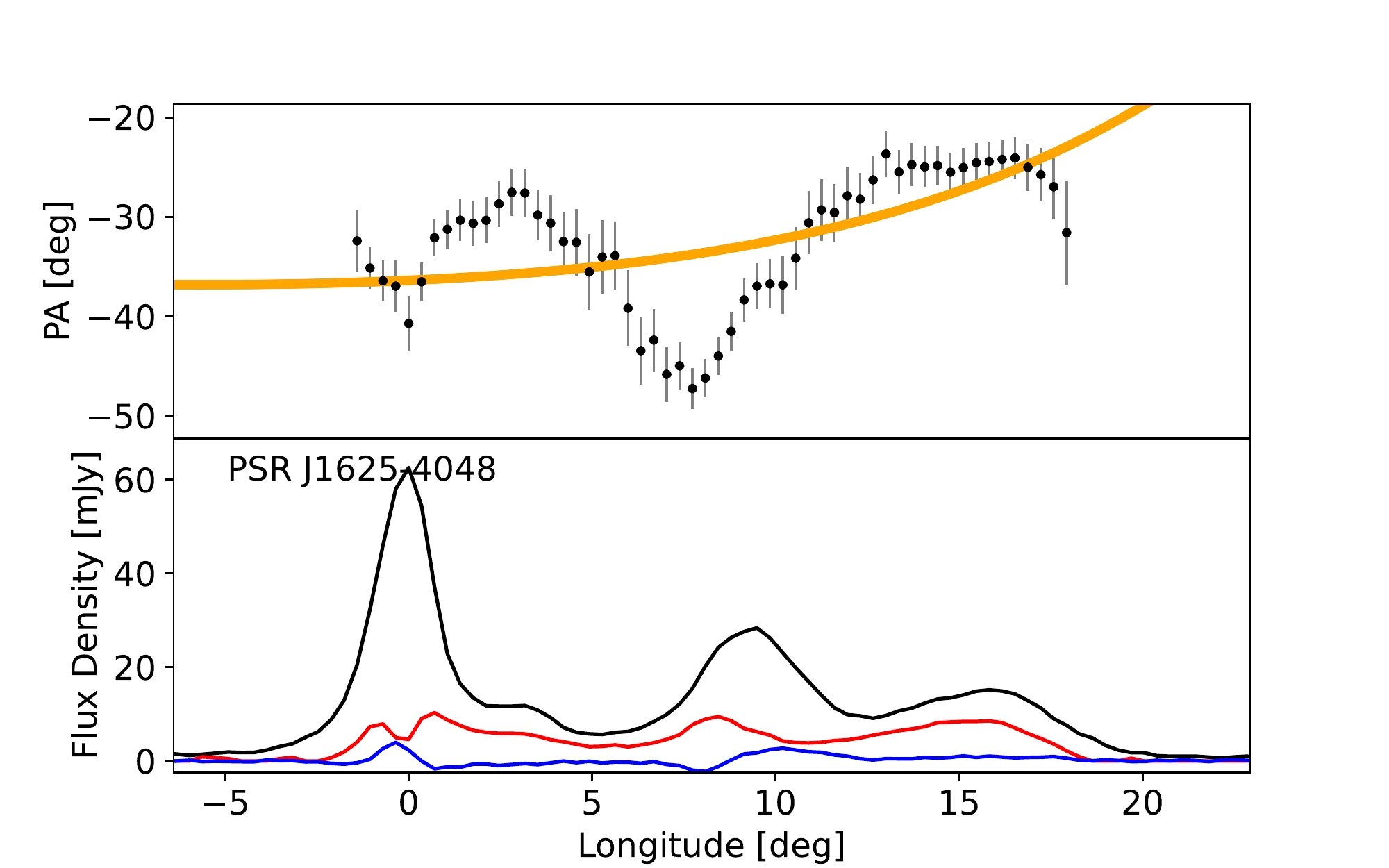} &
\includegraphics[width=9.5cm,angle=0]{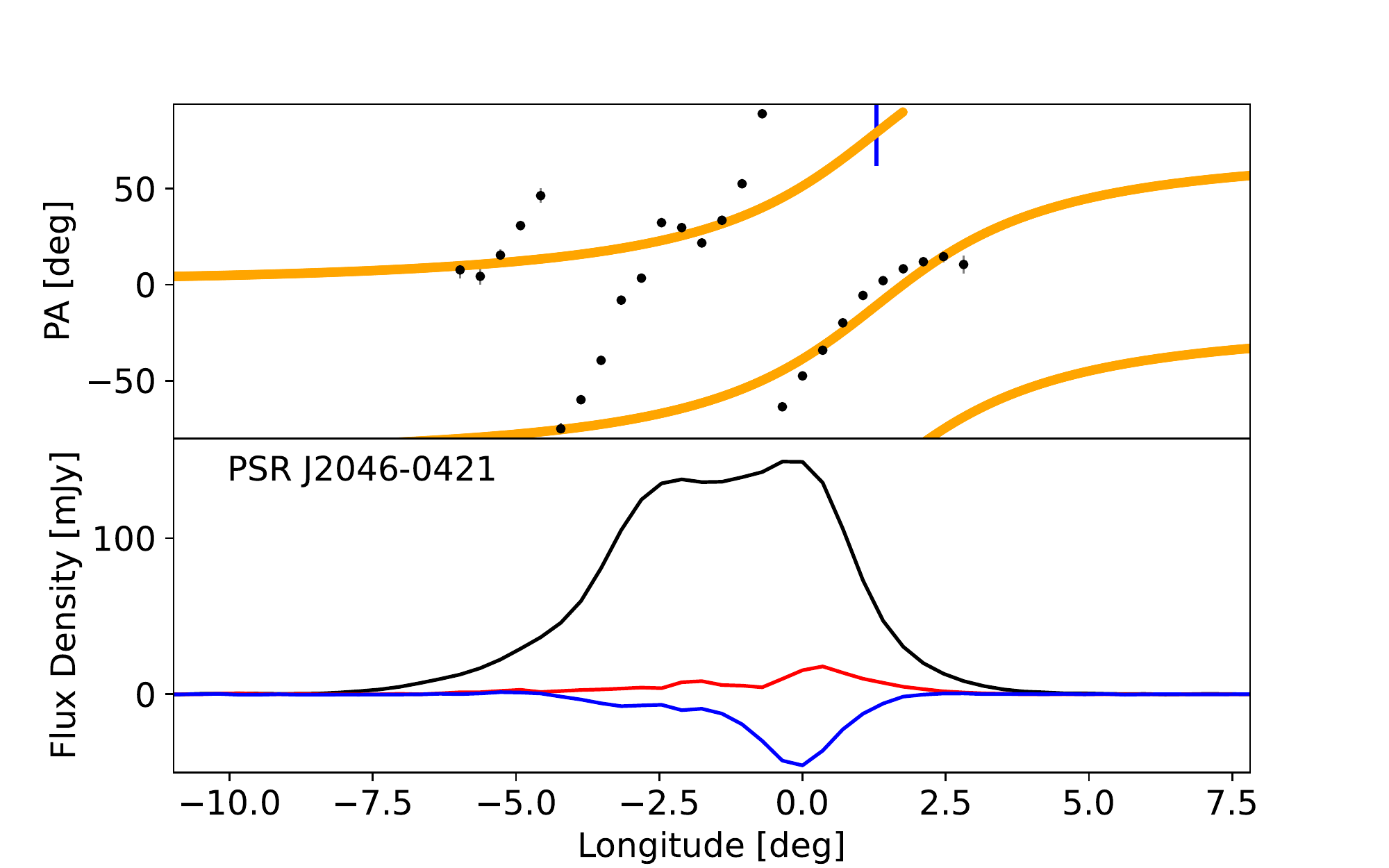} \\
\includegraphics[width=9.5cm,angle=0]{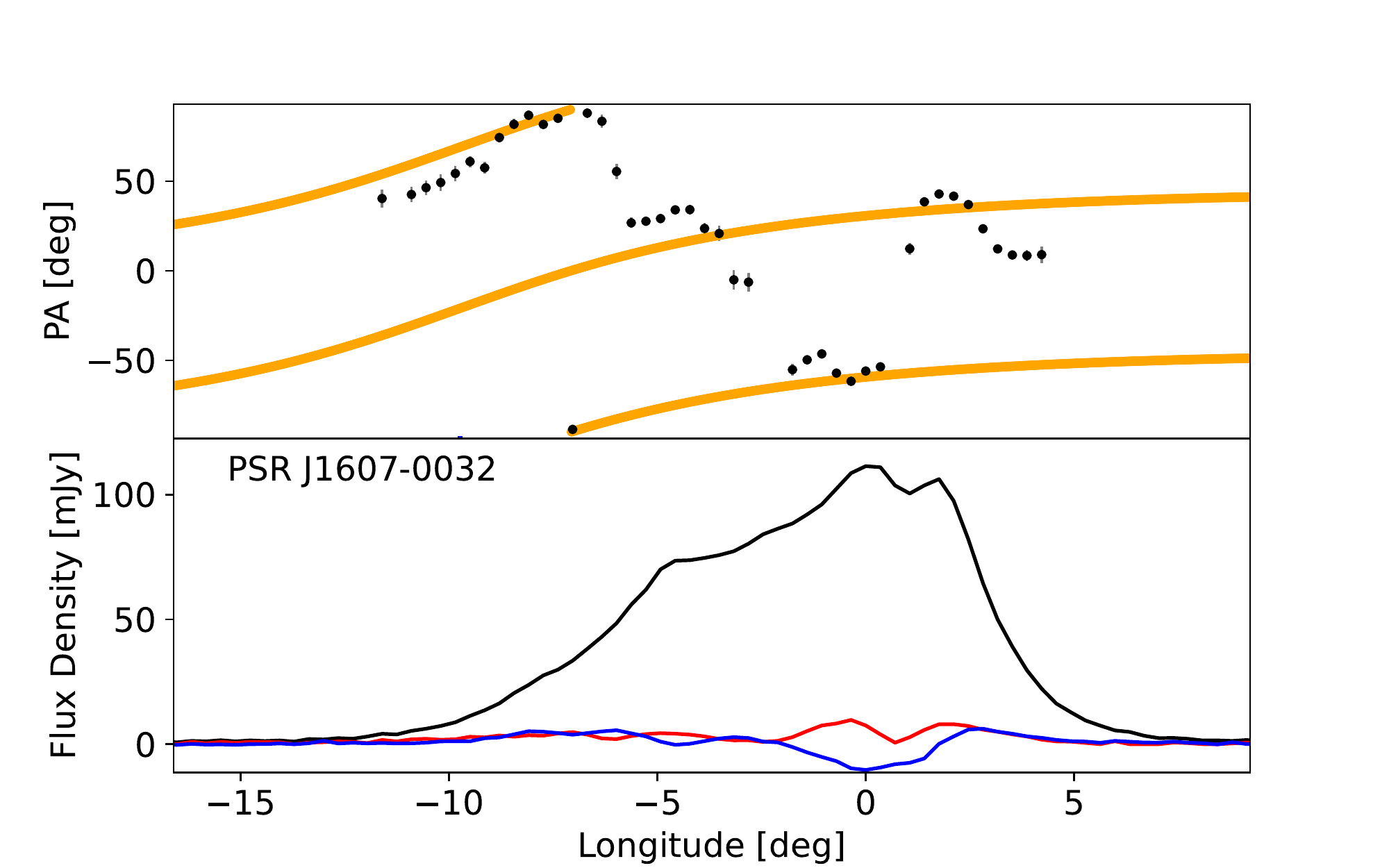} &
\includegraphics[width=9.5cm,angle=0]{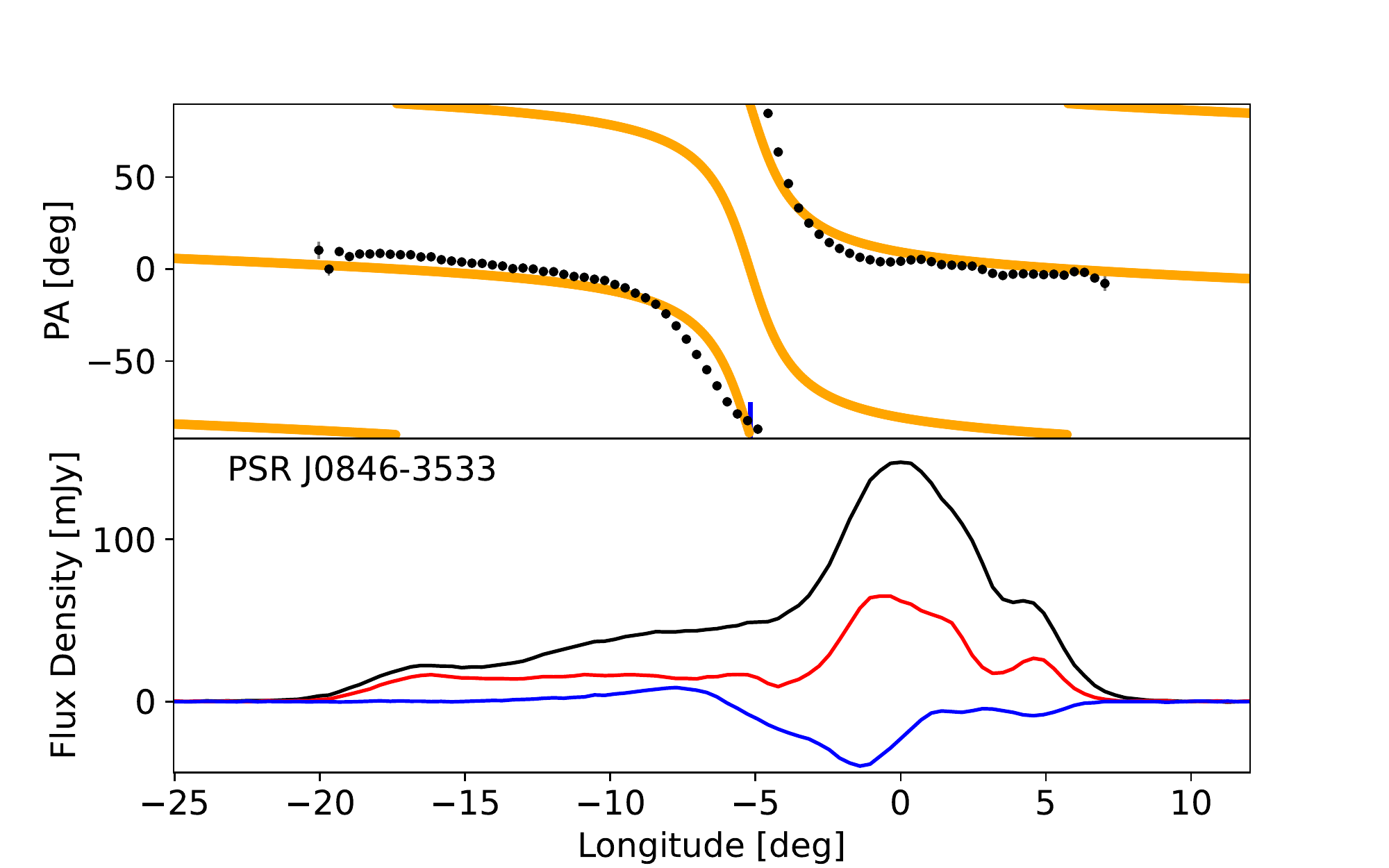} \\
\end{tabular}
\end{center}
\caption{Examples of pulsars in the `non RVM' class, for which the RVM fitting returns a high value of $\chi^2$. PSR~J1625--4048 shows an example of a PA which changes direction several times across the profile. PSR~J2046--0421 shows a very large PA swing across the profile. PSR~J1607--0032 shows high levels of circular polarization compared to linear polarization. PSR~J0846--3533 shows a PA swing with only minor deviations from an RVM-like curve.}
\label{figcrazy}
\end{figure*}

\section{Observations and calibration}
Observations were conducted as part of the Thousand Pulsar Array (TPA) program, itself part of the larger MeerTime project on the MeerKAT telescope. Details of the project as a whole and the observational setup can be found in \citet{mtime} and \citet{tpa20}. In brief, we used the observational band from 896 to 1671~MHz with 928 frequency channels. The data are folded into sub-integrations each of length 8~s for the duration of the observation and there are 1024 phase bins per pulse period. Polarization calibration is carried out using the procedures described in detail in \citet{sjk+21}. Flux calibration is carried out as described in \citet{poss+22}. All these operations are performed by the processing pipeline \textsc{meerpipe}\footnote{https://github.com/aparthas3112/meerpipe} which produces RFI-excised, polarization- and flux-calibrated output products in \textsc{psrfits} format \citep{hvm04}. A more complete description of the pipeline is reported in \citet{pbs21}. In this paper we use the same data set described in detail in \citet{poss+22}.

\section{RVM fitting}
The input data into the RVM fitter was a set of PAs and associated values of longitude, $\phi$, taken from the set of pulsars presented in \citet{poss+22}. The PAs were derived from Stokes Q and U, and were only determined when the linear polarization was more than $5\sigma$ significant. Error bars on the PAs were obtained from the value of the linear polarization compared to the noise using the procedure from \citet{ew01}. The peak of the profile was generally chosen as the zero point of longitude. RVM fitting was carried out using a least-squares procedure implemented in {\sc python}. A total of 85 trials in $\alpha$ were used ranging from 5\degr\ to 175\degr. In $\beta$, 40 trials were used, initially between $\pm$20\degr, a range which could be refined on subsequent runs. For each $\alpha$,$\beta$ pair, the {\sc scipy} routine {\sc least\_squares} was used to determine the values of $\phi_0$ and PA$_0$ which minimized $\chi^2$ and the minimum $\chi^2$ was recorded.  We note that PAs can only lie between $-90$\degr\ and $+90$\degr\ and hence the phase-wraps need to be taken into account. In addition, radio emission can appear in two orthogonal modes, and these so-called orthogonal mode transitions must also be built into the fitting routine.

In this work we use the so-called `observers convention' which defines position angles as increasing counter-clockwise on the sky. This means that when the slope of the PA traverse is positive then the sign of $\beta$ is negative. Also, when $\alpha < 90$\degr\ and $\beta > 0$\degr\ then we observe an {\bf outer} line of sight (i.e. away from the spin axis and towards the equator). However, for $\alpha > 90$\degr\ and $\beta > 0$\degr\ we observe an {\bf inner} line of sight (i.e. nearer to the spin axis). Therefore the sign of $\beta$ alone does not reveal whether the sight line is inner or outer. For a detailed discussion see \citet{ew01}.

\section{RVM fitting results}
The TPA census consists of 1267 pulsars and is fully described in \citet{poss+22}. For the 21 interpulse pulsars in our sample, we performed independent fits on the main and interpulse components where sufficient PA points were available. Not all of the pulsars are amenable to RVM fitting. We required at least 10 PA points across the profile to proceed with the fit. This effectively removes pulsars for which the signal to noise ratio is low and pulsars which are bright but have very low level of linear polarization. In addition we removed pulsars which are scattered at our observing frequency as recorded in \citet{okp+21}. Scattering has the effect of causing distortions in the PA swing due to the convolution of the Stokes Q and U profiles with an exponential function \citep{kar09}. These criteria led to a removal of 434 pulsars from the sample, listed in Table~\ref{tab:no}. This leaves a grand total of 854 RVM fit attempts.

\begin{figure}
\begin{center}
\begin{tabular}{ccc}
\includegraphics[width=8cm,angle=0]{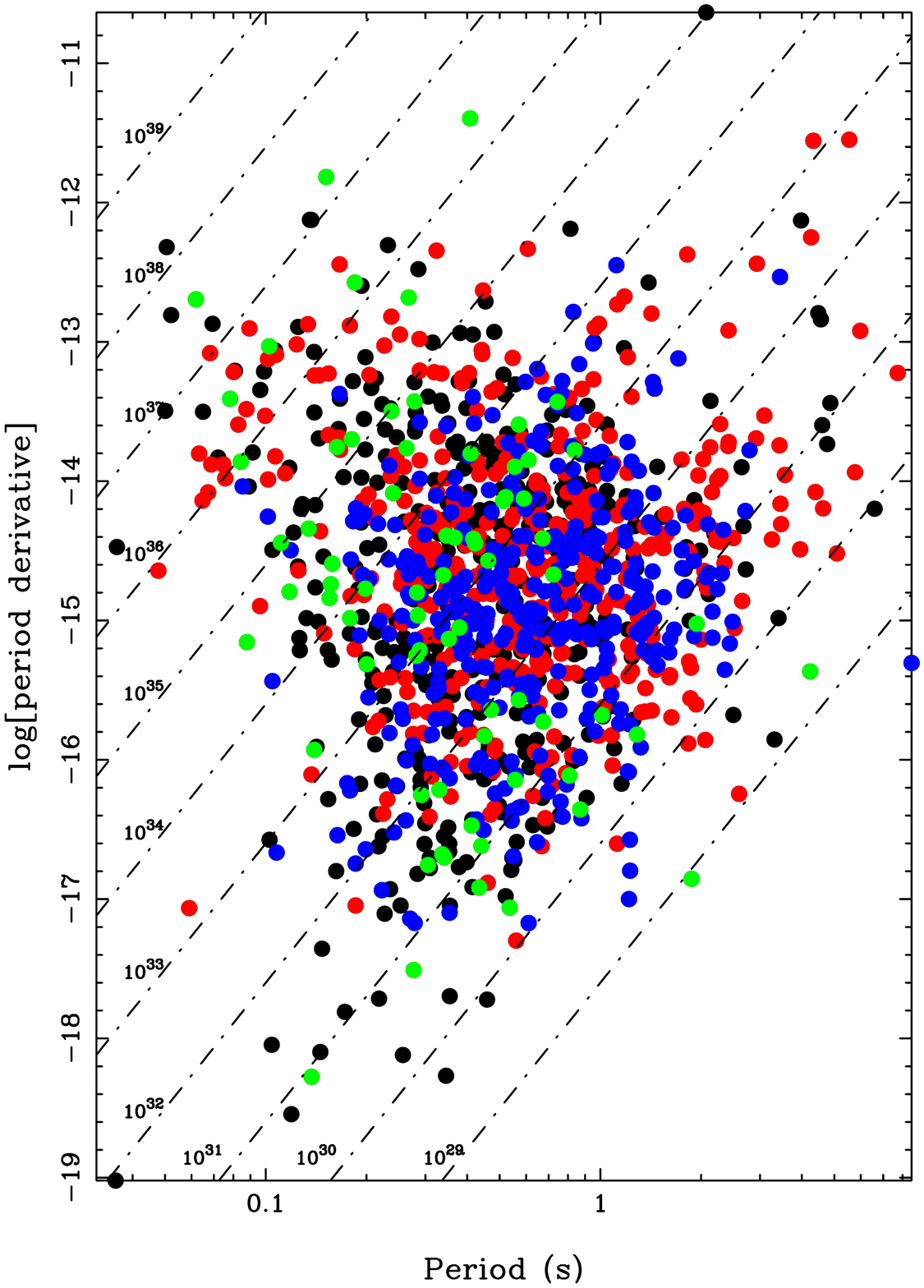}\\
\end{tabular}
\end{center}
\caption{The period-period derivative plane of the pulsars in the sample (black). The RVM class are shown in red, the `flat' class in green and the non-RVM in blue). Lines of constant $\dot{E}$ are shown with dash-dots.}
\label{figppdot}
\end{figure}

There are three main classes of output from the fitter as detailed below, examples of which are given in Figures~\ref{figgood}, \ref{figflat} and \ref{figcrazy}. In Figure~\ref{figgood} the left-hand panels show the pulsar profile in total intensity, linear and circular polarization as a function of longitude, along with the position angle of the linear polarization. The right-hand panel shows an image of $\chi^2$ values on the $\alpha-\beta$ plane. Lack of colour denotes $\chi^2 > 10$. In addition, contours of constant emission height are shown based on knowledge of $W_{10}$ via equations~\ref{height} and \ref{rho}. Meanwhile, Figures~\ref{figflat} and \ref{figcrazy} show the profiles and the PA traverse only. The numbers in each class are shown in the second column of Table~\ref{tabtype} and a visual representation of the results on the $P-\dot{P}$ plane is shown in Figure~\ref{figppdot}. A machine-readable ascii table with the full results is provided in the on-line material.

\subsection{Class `RVM'}
The pulsars in this class are those for which the RVM fitter returns a low value of reduced $\chi^2$, typically $<2$. Information on these pulsars is given in Table~\ref{tab:rvm}. In this class of pulsar, the PA is well-behaved over the entire profile, after orthogonal mode jumps are taken into account (see for example PSR~J1110--5637 in Figure~\ref{figgood}). The large majority of pulsars in this category show the classic `banana-shaped' $\chi^2$ contour plot on the $\alpha$-$\beta$ plane. This implies that $\alpha$ is unconstrained, the sign of $\beta$ is determined, that $|\beta|$ has a maximum allowed value and that there exists a good $\chi^2$ fit for a particular $\alpha,\beta$ pair.

The `banana' can be thin either when the fit is excellent or when there is sufficient change of slope across the PA swing to sufficiently constrain $\beta$, as with PSR~J1048--5832 in Figure~\ref{figgood}. The `banana' becomes thick for fits which are less good or when the slope has only changed marginally across the profile. An example is PSR~J1722--4400 in Figure~\ref{figgood}.
In rare cases, the `banana' can be resolved into a well-constrained part of the $\alpha$-$\beta$ plane. This happens only when the longitudonal extent of the profile is very wide or the pulsar has an interpulse. An example of the former is PSR~J1842--0359 in Figure~\ref{figgood}.
It should also be noted that the values of PA$_0$ and $\phi_0$ are (largely) independent of $\alpha$ and $\beta$. This is because the inflection point of the PA swing and hence the value of the PA at that location are relatively model-independent.

Finally, it can be seen that the lines of constant emission height make an almost orthogonal cut through the `banana'. If therefore the emission height is known, this provides a tight constraint on the $\alpha$,$\beta$ pair. Conversely if the geometry is well determined then the emission height can be derived.

\subsection{Class `flat'}
This class contains pulsars with a flat swing of PA across the profile, with a maximum swing of less than 2 degrees per degree and/or very little change in the overall slope as a function of longitude. For these pulsars, we cannot constrain either $\alpha$ or $\beta$, although the sign of $\beta$ can at least be determined. The pulsars in this class are listed in Table~\ref{tab:flat}, with four examples shown in Figure~\ref{figflat}.

\subsection{Class `non RVM'}
Pulsars in this class contain PA swings which are forbidden by the RVM and are listed in Table~\ref{tab:bad}. For example, some pulsars show several changes of direction in PA across the profile (see PSR~J1625--4048 in Figure~\ref{figcrazy}), while others show very large overall swings in PA (see PSR~J2046--0421 in Figure~\ref{figcrazy}). The fitting process returns high values of $\chi^2$ for these pulsars and no sensible constraints on the geometry can be obtained. However, for about 25\% of the pulsars in this class, the PA swing looks broadly RVM-like, but with sufficient deviations to render the $\chi^2$ high. An example is PSR~J0846--3533 in Figure~\ref{figcrazy}.

\section{Discussion}
\begin{figure}
\begin{center}
\begin{tabular}{c}
\includegraphics[width=8cm,angle=0]{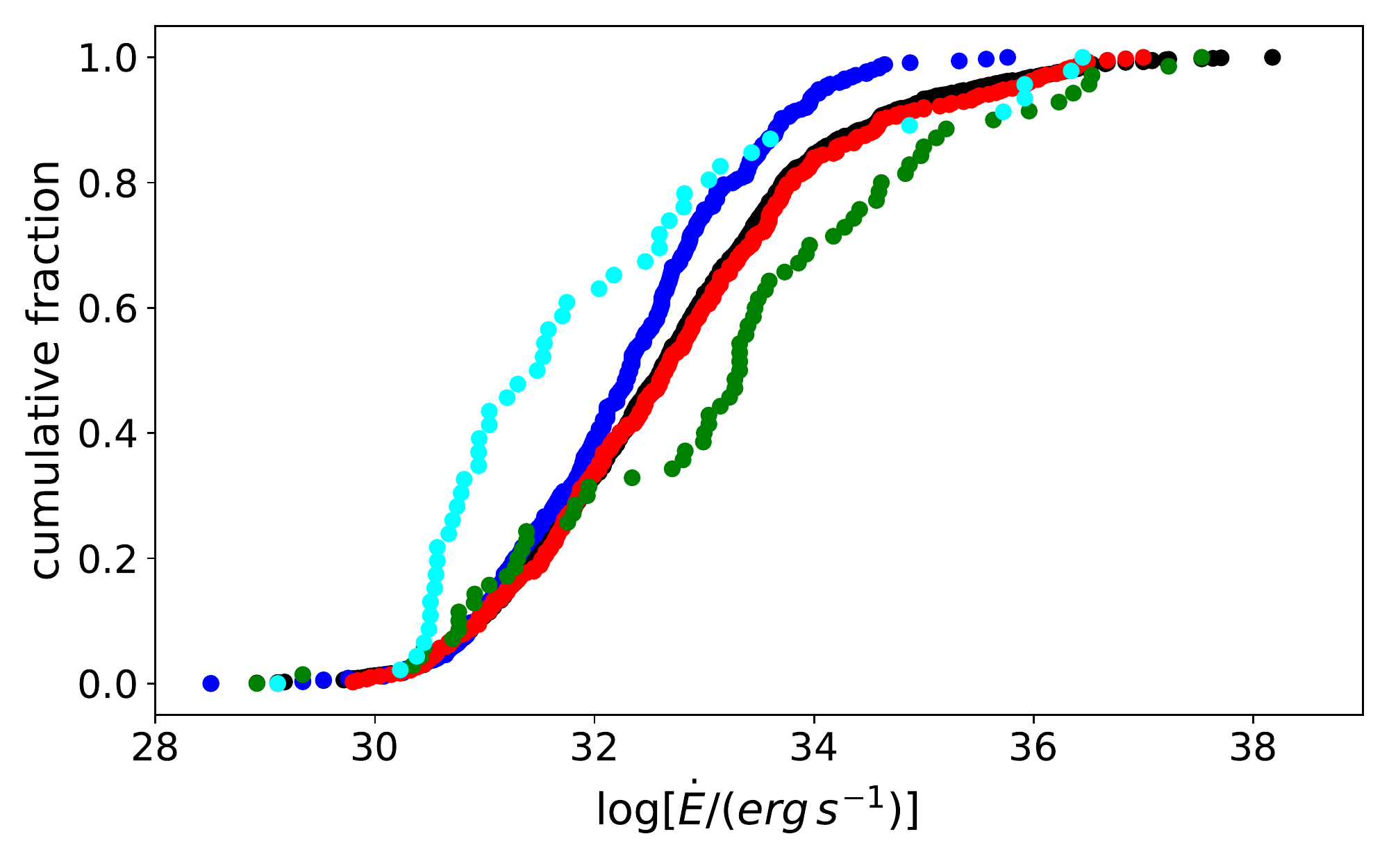} \\
\end{tabular}
\end{center}
\caption{Cumulative $\dot{E}$ distributions for the whole sample (black), the RVM class (red), the non-RVM class (blue), the flat class (green) and pulsars with low values of $\alpha$ (cyan).}
\label{figedotclass}
\end{figure}

\subsection{RVM versus non-RVM pulsars}
Table~\ref{tabtype} indicates that for the sample of 854 pulsars, some 60\% of them have PA swings which appear RVM-like and 40\% do not. What are the differences between these two classes? Figure~\ref{figedotclass} shows the distribution of $\dot{E}$ for the entire sample, the RVM class, the flat class and non-RVM class. It can be seen that the RVM class has a distribution almost identical to that of the total sample whereas the flat class of pulsars is biased towards high $\dot{E}$ and the non-RVM pulsars show a bias towards low values of $\dot{E}$. Indeed only 8 out of 77 pulsars with $\dot{E} > 10^{34.5}$~erg\,s$^{-1}$ are in the non-RVM class. In many ways this is not a surprise. Pulsars with high $\dot{E}$ have previously been shown to have `simpler' profiles with high degrees of linear polarization and unbroken PA swings \citep{jw06,rwj15b}. This may arise because their emission heights are relatively high \citep{kj07} thus minimising the distortion of the polarization properties on the traverse through the magnetosphere. In contrast, the pulsars with low $\dot{E}$ often have profiles which are a blend of multiple components (see Figure~\ref{figcrazy}), possibly arising over a large range of emission heights, thereby distorting the PA swing (e.g. \citealt{cr12,dyks17}).

From visual inspection it is also apparent that pulsars in the non-RVM class have significantly more circular polarization (compared to linear polarization) than pulsars with RVM-like swings. To quantify this we determined how many pulsars have $|V| > L$ in more than 5 phase-bins across the profile, which we denote as $V_{\rm h}$ in Table~\ref{tabtype}. There are 288 such pulsars, 72\% of which occur in the non-RVM class. Expressed another way, pulsars denoted as $V_{\rm h}$ occur 16\% of the time in the RVM class and 59\% in the non-RVM class. It therefore appears as a general, if not hard-and-fast rule, that distortions of the PA swing occur simultaneously with a high degree of circular polarization (see for example PSR~J1607--0032 in Figure~\ref{figcrazy}). Generation of circular polarization has long been a thorny problem in pulsar astronomy \citep{rr90,km98}. In the simplest picture where two linearly polarized orthogonal modes interact incoherently, neither circular polarization nor distortion of the RVM curve can happen. Recently, \citet{dyks19,dyks20} pointed out that coherent addition of the modes with some (non-zero) phase lag not only generates circular polarization, but can also reproduce many of the peculiar PA traverses seen in single pulses. Motivated by this, \citet{osw+22} developed a framework for comparing the polarization properties of a large sample of pulsars to draw conclusions about the ratio of incoherent to coherent mixing (Oswald et al. In Prep.).

In the data presented here we also see manifestations of the same effect. In pulsars with an observed high fraction of linearly polarization, one mode dominates and any coherent mixing has a negligible effect on the PA swing. These tend to be the high $\dot{E}$ pulsars. In pulsars with lower $\dot{E}$, the modes are more evenly matched in amplitude. As a result the linear polarization is low and coherent mixing can then produce significant circular polarization accompanied by distortions of the PA swing as observed. In a future paper we aim to take these ideas further by considering what mixing parameters are required to `fix' the PA swing back to the geometrical RVM curve.

\subsection{Flat PA swings}
The pulsars with flat PA swings tend to be at high $\dot{E}$ as shown in Figure~\ref{figedotclass}. This class of pulsars was identified already in \citet{jw06}. They have relatively short periods and hence large polar caps. This makes for a large beam and if $\alpha$ is low and the line of sight cuts the beam at high $|\beta|$ then the result is a wide profile with a flat PA swing (e.g. PSR~J1524--5625, see Figure~\ref{figflat}). Alternatively the emission height could be rather large, this would result in the PA swing leading the profile by a large amount (equation~\ref{eqn:dp}) meaning that we would miss the inflection point entirely.

However, there exists an interesting sub-category of pulsars with flat PA swings, low $\dot{E}$ and relatively high linear polarization fraction. Some of these pulsars have profiles with very large widths such as PSRs~J1232--4742 (Figure~\ref{figflat}) and J1842+1332; presumably these are nearly aligned rotators where the line of sight remains within the beam. This is reminiscent of the profiles of radio-loud magnetars such as XTE J1810--197 \citep{crj+07} and PSR~J1622--4950 \citep{lbb+12}. Others such as PSRs~J0108--1431 and J1511--5414 (both shown in Figure~\ref{figflat}) have narrow profiles, either because the line-of-sight just grazes the beam edge or because the emission is from an isolated patch far from the magnetic axis on the leading or trailing edges.

\subsection{Relativistic effects}
More than 30 years ago, \citet{bcw91} pointed out that the inflection point of the RVM curve ($\phi_0$) should trail the centroid of the intensity profile by an amount which depends linearly on the rotation period and the emission height. For the pulsars in our sample we define the centroid as lying half-way between the 10\% intensity levels of the profile, which we label $\phi_m$. Then, $\Delta_\phi$ is simply $\phi_0 - \phi_m$ which can be related to the physical parameters via Equation~\ref{eqn:dp}.

\begin{figure}
\begin{center}
\begin{tabular}{c}
\includegraphics[width=9cm,angle=0]{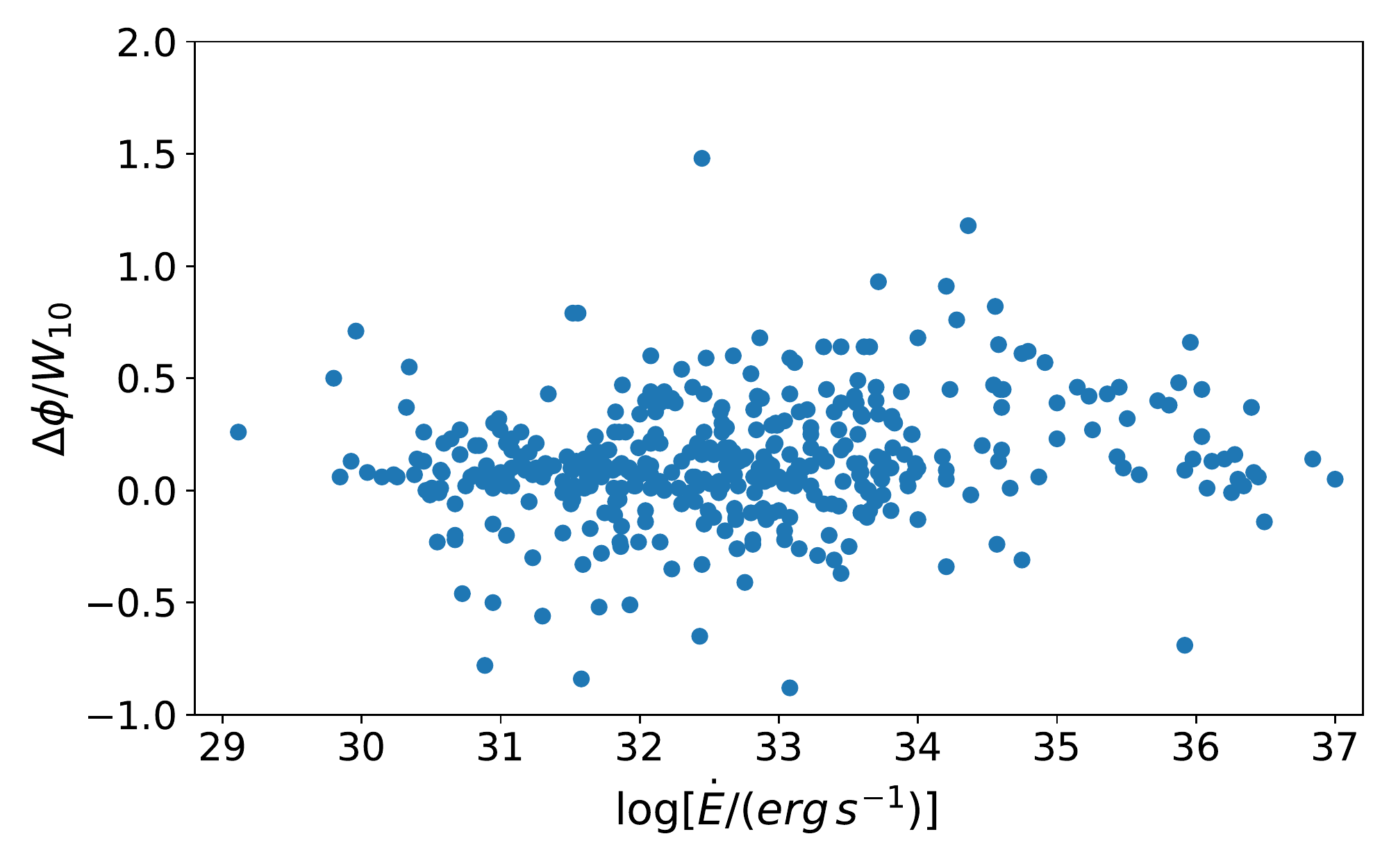} \\
\end{tabular}
\end{center}
\caption{Spin-down energy loss-rate $\dot{E}$ versus $\Delta\phi / W_{10}$ for a sample of 411 pulsars. See text for details.}
\label{figphi}
\end{figure}

We find that $\Delta_\phi$ is within 1\degr\ of zero in 20\% of cases, $\Delta_\phi > 1$\degr\ in 62\% of pulsars and $\Delta_\phi < -1$\degr\ for 18\% of the population (see Table~\ref{tabtype}). This strong bias towards positive values indicates the \citet{bcw91} effect at work. Figure~\ref{figphi} shows $\Delta_\phi / W$ versus $\dot{E}$. Here we see that virtually all pulsars with $\dot{E} > 10^{34}$~erg\,s$^{-1}$ have $\Delta_\phi/W > 0$\degr (53 out of 61 cases). This is likely due to a combination of short $P$ and high heights (cf Equation~\ref{eqn:dp}). For pulsars with $\dot{E} < 10^{34}$~erg\,s$^{-1}$ there remains a strong bias towards positive values of $\Delta_\phi/W$ (262 out of 350 cases). Some of the pulsars with negative $\Delta_\phi$ are so-called `partial cones' which are dealt with in more detail in subsection 5.6.

\subsection{Emission Heights}
\begin{figure}
\begin{center}
\begin{tabular}{c}
\includegraphics[width=9cm,angle=0]{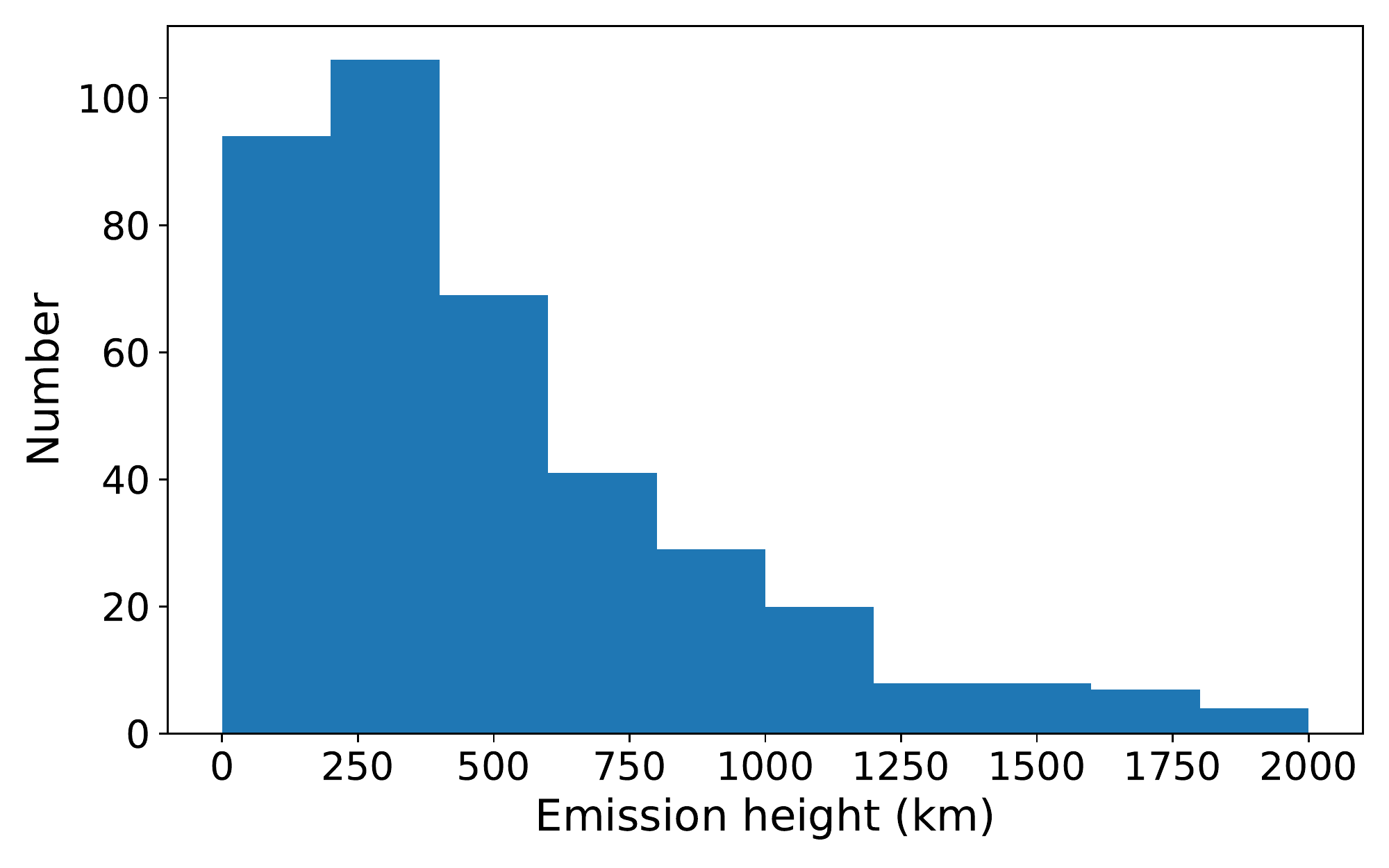} \\
\end{tabular}
\end{center}
\caption{Histogram of the emission heights of 386 pulsars with good RVM fits under the assumptions that $\alpha=90$\degr and that the beams are filled. A further 35 pulsars with $h_{\rm max} > 2000$~km are not shown.}
\label{fighmax}
\end{figure}

We can compute a limit on the emission height for the pulsars in the RVM class in the following (geometric) way. If we assume that the beams are filled and that $\alpha=90$\degr\ for all pulsars, an upper limit on the emission height is given by Equations~\ref{height} and \ref{rho}. The resultant histogram of heights computed in this way is shown in Figure~\ref{fighmax}. From this figure we see that 35\% of the pulsars have $h_{\rm max} < 300$~km and 77\% have $h_{\rm max} < 1000$~km. This provides strong evidence that the radio emission from pulsars arises from low altitude, and does not depend greatly on the spin period \citep{jk19,pjk+21}. In addition we also note that pulsars with apparent $h_{\rm max} > 1000$~km have predominantly low values of $\dot{E}$. Given that $\alpha$ evolves with time, these pulsars likely have $\alpha$ significantly less than 90\degr\ and as a result lower emission heights than $h_{\rm max}$. 

We can also compute emission heights using the relativistic effects described in the previous subsection. Knowledge of $\Delta\phi_0$ can be converted to a height using Equation~\ref{eqn:dp}. We find that, at most, only 10\% of the pulsars can have $h_{\rm max} > 1000$~km when measured in this way. In summary therefore, the geometric approach and the relativistic approach give reasonable agreement and we surmise that virtually all the pulsars in this sample have $h_{\rm max} < 1000$~km, independent of spin period.

\subsubsection{Pulsars with high $\dot{E}$}
There are 55 pulsars in the sample with $\dot{E}>10^{35}$~erg\,s$^{-1}$, only 5 of which are in the non-RVM class. A high fraction of these pulsars are seen at X-ray and/or $\gamma$-ray energies. In the outer-gap or two-pole caustic model of $\gamma$-ray emission \citep{dr03,wr11} the expectation is that these pulsars are predominantly seen at high values of $\alpha$. In addition if $\alpha$ is randomly distributed at birth, then geometrical considerations again imply that high values of $\alpha$ should dominate in the observed population. However, a comprehensive study of these pulsars by \citet{rwj15,rwj15b} showed that either the $\alpha$ distribution is skewed towards low values or emission must arise from an open field-line region which is larger than conventionally defined from equation~\ref{height}.

Recent work on $\gamma$-ray emission models \citep{pm21} have shown that high values of $\alpha$ may not be necessary \citep{dpm22}. A study of interpulse pulsars with high $\dot{E}$ \citep{jk19b} showed that low emission heights are prevalent and that in many instances the beam is underfilled. In the work presented here, the measured values of $\Delta\phi$ also generally prefer low emission heights (below 500~km). These low emission heights coupled with large widths again imply low values of $\alpha$ via equation~\ref{rho}. We are therefore forced into the same conclusion as \citet{rwj15} that either $\alpha$ is not randomly distributed or that emission occurs over a larger region than the classic polar cap extent. Indeed in a force-free magnetosphere the closed field-line region does not reach the light cylinder, thereby extending the region bounded by the open field lines (e.g. \citealt{spit06,cra14,pk22}). Additional theoretical support for this idea based on the phenomenology of sub-pulse drifting can be found in \citet{wri22}.

\subsection{Evidence for evolution of $\alpha$ with time}
In the previous section we showed strong evidence that emission heights are less than 1000~km for the majority of the radio pulsar population. If we now {\bf assume} that 1000~km is the maximum allowed emission height, then we can use the width of the profile to provide an upper limit on the value of $\alpha$ by combining equations~\ref{height} and \ref{rho}. As an example, it can be seen that for PSR~J1842--0359 in Figure~\ref{figgood} the maximum value of $\alpha$ is 15\degr. Of the 431 pulsars with good RVM fits, 47 have a maximum allowed $\alpha$ less than 45\degr. The cumulative $\dot{E}$ distribution of these pulsars is shown in cyan on Figure~\ref{figedotclass}. The distribution is dominated by low $\dot{E}$ pulsars, 37 of the 47 have $\dot{E} < 10^{33}$~erg\,s$^{-1}$ and only 5 have $\dot{E} > 10^{35.5}$~erg\,s$^{-1}$. As lower $\dot{E}$ pulsars are older than those with higher $\dot{E}$, this result supports an evolution of the magnetic axis towards the spin axis with time, evidence for which is also apparent in other observational studies \citep{tm98,wj08b,jk17} with theoretical underpinning in \citet{ptl14} and \citet{nbg+20}.

\subsection{Partial Cones?}
\begin{figure*}
\begin{center}
\begin{tabular}{cc}
\includegraphics[width=9cm,angle=0]{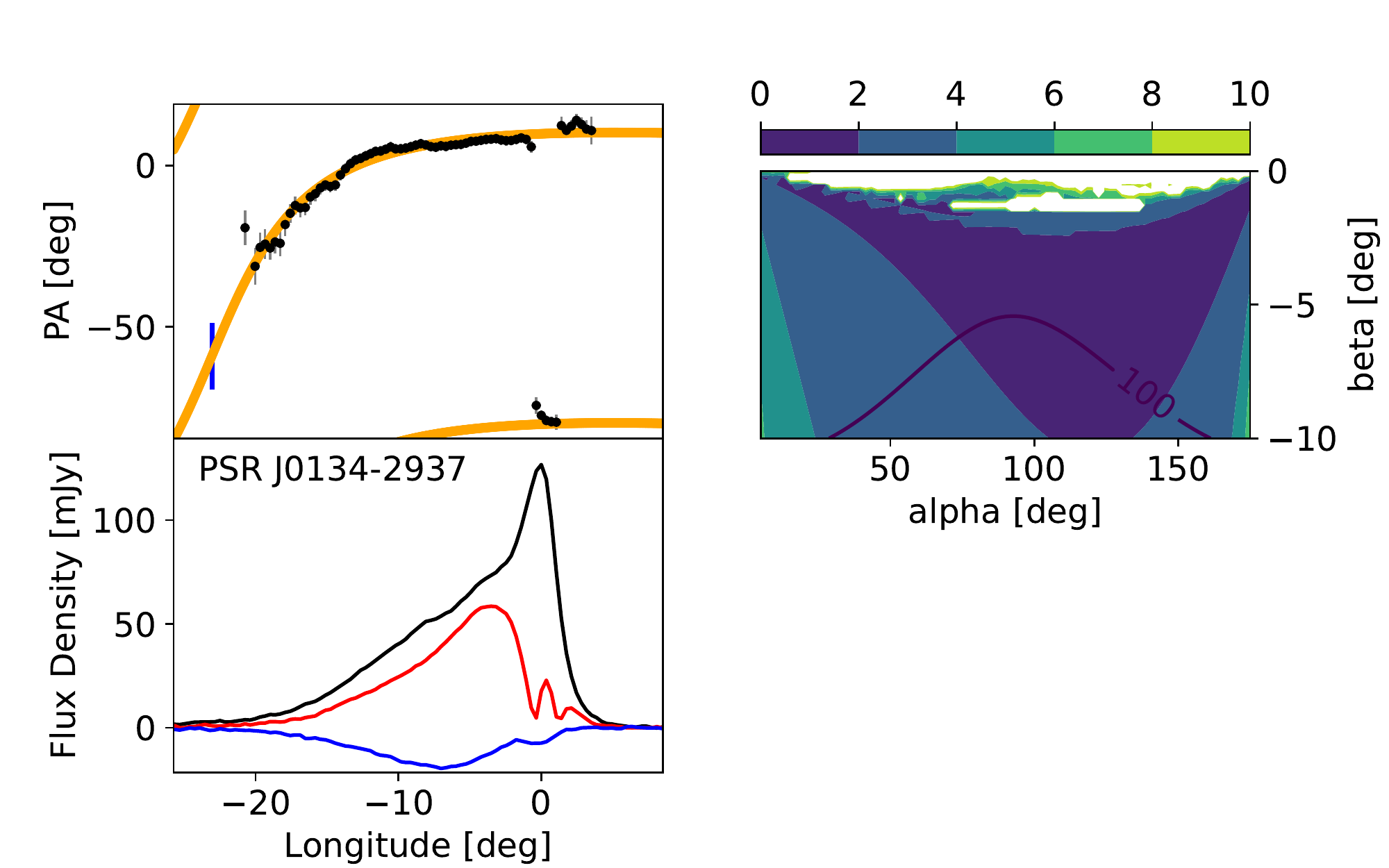} &
\includegraphics[width=9cm,angle=0]{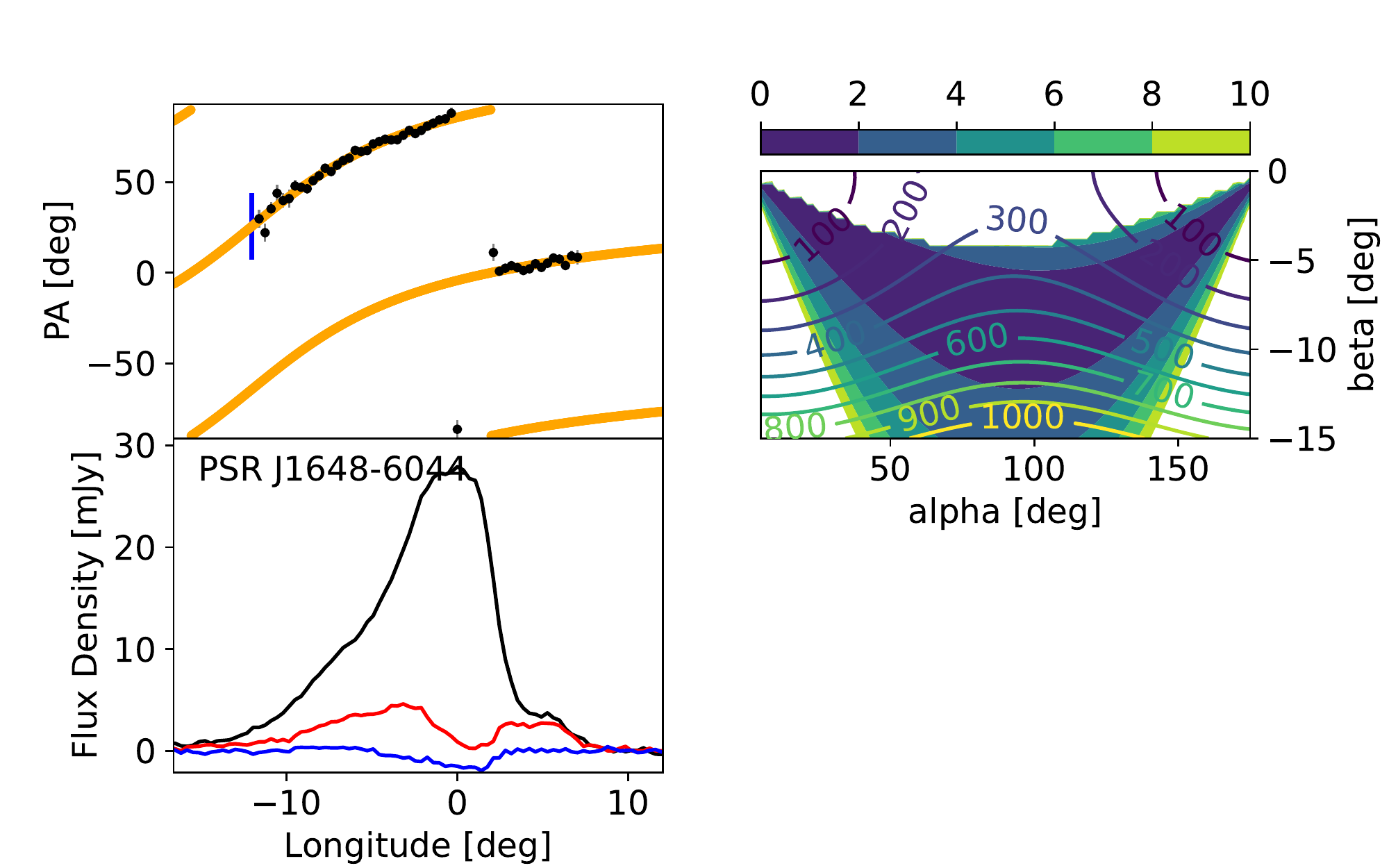} \\
\includegraphics[width=9cm,angle=0]{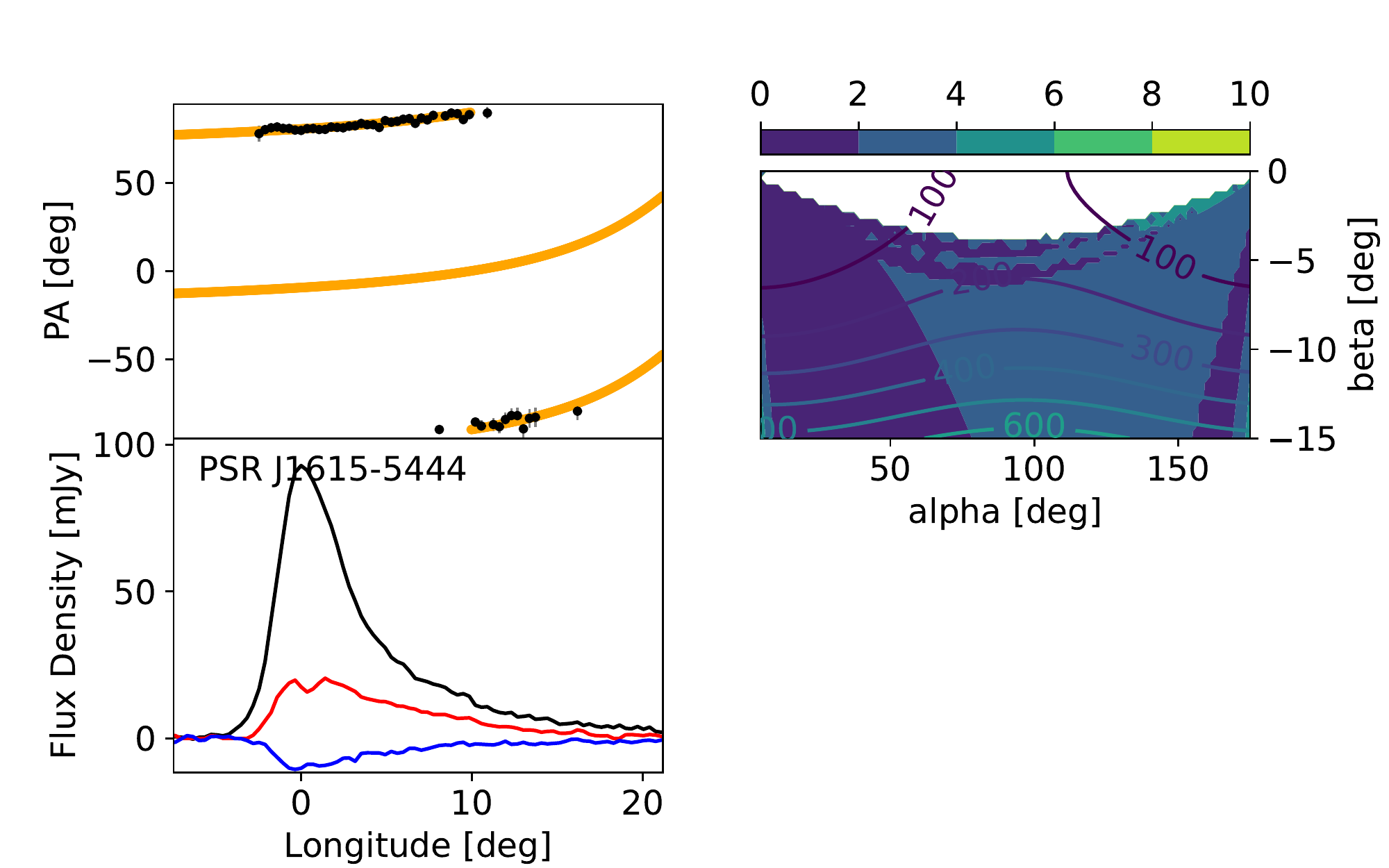} &
\includegraphics[width=9cm,angle=0]{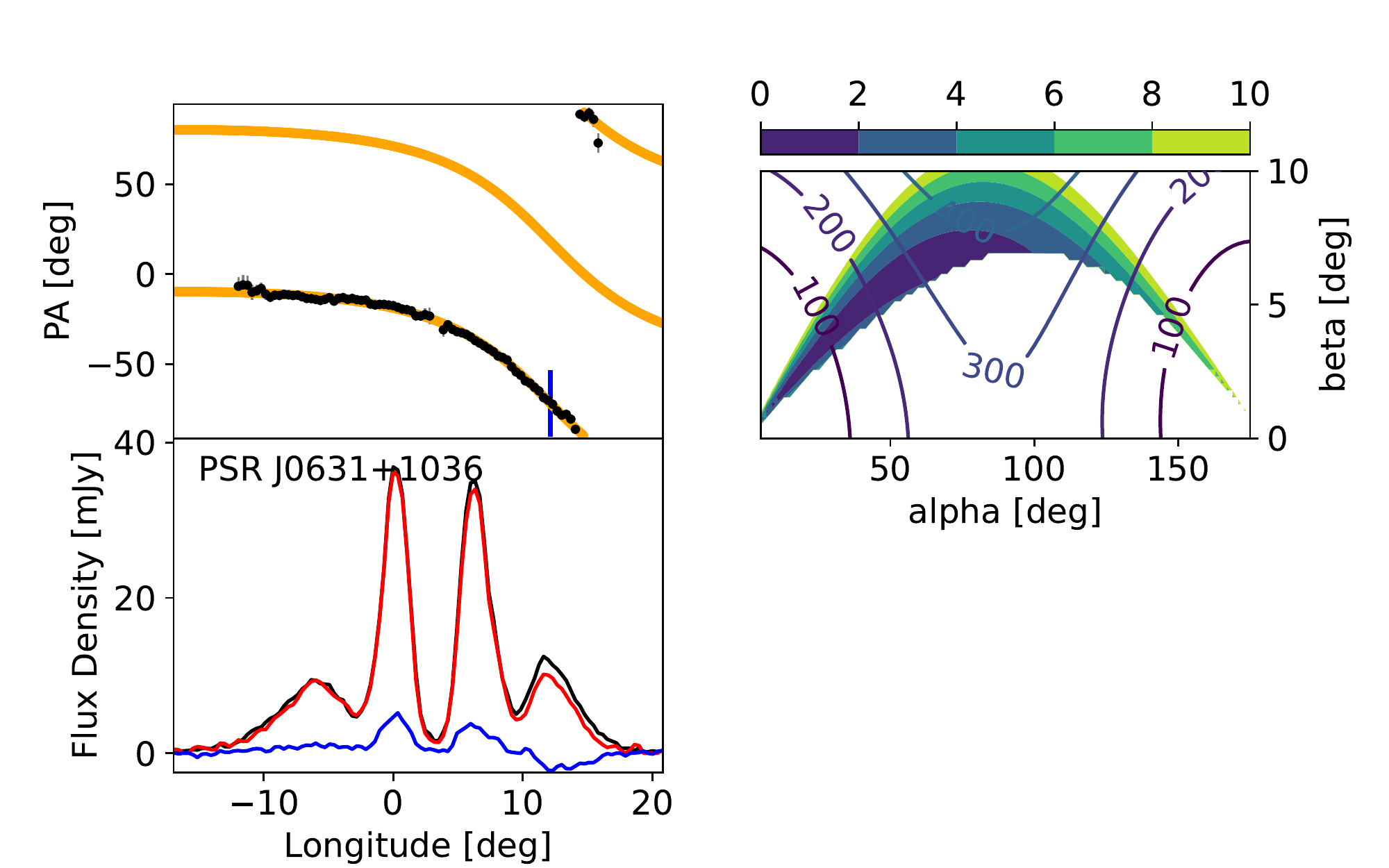} \\
\end{tabular}
\end{center}
\caption{Two examples of trailing partial cone pulsars are PSRs~J0134--2937 (top left) and J1648--6044 (top right). In these two cases $\phi_0$ occurs significantly earlier than the profile midpoint and the profile has a lop-sided look. An example of a leading partial cone is PSR~J1615--5444 (bottom left). Finally PSR~J0631+1035 (bottom right) is an example of a symmetrical profile with $\phi_0$ shifted later due to relativistic effects.}
\label{figpartial}
\end{figure*}
If the pulsar beam consists of nested cones, as in \citet{ran90}, there may be instances where the leading or trailing part of the cone is below the detection threshold and this would manifest itself as a `partial cone' in the nomenclature of \citet{lm88}. In the patchy beam model similarly there may be occasions in which only the leading or trailing part of the beam is illuminated. In the study of 200 pulsars by \citet{lm88} they found up to 50 pulsars (25\%) could be classified as partial cones with a ratio of approximately 2:1 between pulsars where only the leading edge is seen compared to the trailing edge. These results were obtained largely by comparing the inflection point of the PA swing to the midpoint of the pulse profile. A few years later, \citet{bcw91} pointed out that, due to finite emission heights, it was expected that the PA swing would be delayed compared to the pulse profile midpoint. In retrospect therefore, we see that many (perhaps all) of the leading partial cones in \citet{lm88} are mis-classified once these relativistic effects are taking into account. Furthermore, more sensitive observations carried out on the \citet{lm88} sample by \citet{mr11} concluded that there were no good examples of the inflection point leading the profile midpoint and that trailing partial cones were not seen.

We largely concur that none of the pulsars identified in \citet{lm88} are convincing as trailing partial cones. However, we find several striking examples of this amongst our sample from pulsars discovered post-1988. The most prominent examples are PSRs~J0134--2937, J1648--6044 (both shown in Figure~\ref{figpartial}) and J2234+2114  with perhaps 2-3 other less convincing cases. In total therefore the trailing partial cones form only a very small minority ($<5$\%) of the overall population. For pulsars which have $\phi_0$ later than the profile midpoint, the majority can be put down to aberration and retardation effects. An example is PSR~J0631+1036 (Figure~\ref{figpartial}) which has a beautifully symmetric four-component profile yet $\phi_0$ occurs significantly later than the symmetric midpoint. Perhaps the two best examples of potential leading partial cones are PSRs J1615--5444 (Figure~\ref{figpartial}) and J1847--0605 but again the fraction of this type of pulsar in the data is very low.

This should not be taken to mean that all pulsar profiles show symmetrical emission patterns about both sides of the magnetic pole. That this is not the case is seen most obviously in the Vela pulsar which clearly lacks emission in the trailing part of the cone. Some of the pulsars with interpulse emission also have highly non-symmetric emission patterns (see \citealt{jk19b}).

\subsection{Orthogonal mode jumps}
\begin{figure}
\begin{center}
\begin{tabular}{c}
\includegraphics[width=9cm,angle=0]{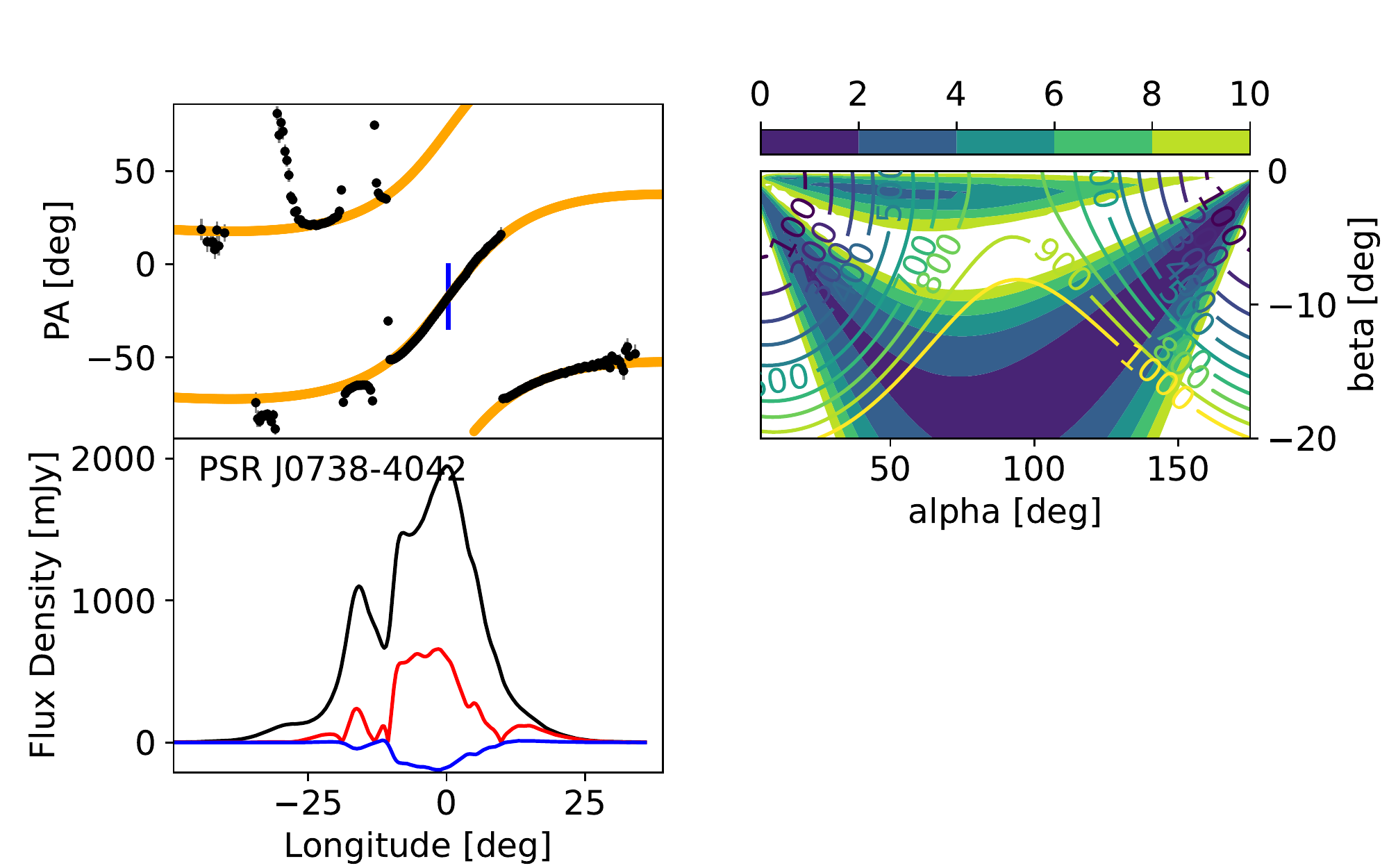} \\
\end{tabular}
\end{center}
\caption{PSR~J0738--4042, an example of a pulsar with orthogonal mode jumps which are not instantaneous but resolved in phase. This causes the RVM fitter to return a high value of $\chi^2$.}
\label{fig0738}
\end{figure}

The presence of two orthogonally polarized emission modes in pulsar profiles has been known since the 1970s \citep{brc76}. If the dominant emission mode changes across the profile, a 90\degr\ jump in the PA swing is observed (e.g. PSR~J1110--5637 in Figure~\ref{figgood}). If we examine the profiles of all the pulsars in the `RVM' and `flat' class, we find that at least one orthogonal jump occurs in 139 out of 502 cases (28\%). We see that the prevalence of orthogonal jumps is lower in the pulsars with a high $\dot{E}$, with only 8 out of 81 pulsars with $\dot{E} > 10^{34}$~erg\,s$^{-1}$ showing this effect. We note that the fraction of pulsars showing an orthogonal jump should be considered a lower limit. Some of the pulsars in the `non-RVM' category show `resolved' jumps (e.g. PSR~J0738--4042; see Figure~\ref{fig0738}) and some show jumps which differ from 90\degr\ by enough to render the RVM fit poor.

We note that, from these data alone, we cannot tell which is the prevalent dominant mode in the population as a whole. When combined with proper motion information, \citet{jhv+05}, \citet{ran07} and \citet{nkcj12} have shown that in some cases the emission parallel to the magnetic field lines dominate while in yet others the perpendicular emission is the dominant one.

\section{Conclusions}
The Thousand Pulsar Array programme on the MeerKAT telescope has produced a homogenous set of data on more than 1200 pulsars as detailed in \citet{poss+22}. We have used the calibrated polarization profiles to perform RVM fits to 854 pulsars. Our main findings are:
\begin{enumerate}
    \item Some 60\% of pulsars have PA traverses which are amenable to RVM fits.
    \item For the 40\% of pulsars for which the RVM fit fails, a substantial number have a circular polarization fraction which exceeds that of the linear polarization in more than 5 bins across the profile. We surmise that coherent mixing of the linearly polarized modes may be responsible.
    \item From geometrical considerations alone, the emission height of the radio radiation at 1.4~GHz is less than 1000~km and is largely independent of rotation period. Emission heights obtained through relativistic delay of the polarization with respect to the total intensity emission also show that emission heights must be low.
    \item There remains a problem with the high $\dot{E}$ pulsars; either their $\alpha$ distribution is non-random or emission occurs from outside the conventionally defined polar cap.
    \item Under the assumption that emission heights are less than 1000~km across the population then there is strong evidence for alignment of the magnetic and rotation axes.
    \item The dominant emission mode changes across the profile in at least 28\% of the population, a fraction which is lower in the high $\dot{E}$ pulsars.
    \item Clear examples of partial cones are rare in the population as a whole.
\end{enumerate}

\section*{Data Availability}
The complete set of pulsar profiles used in this work is available under the DOI\,10.5281/zenodo.7272361, with all details to be found in \citet{poss+22}.

\section*{Acknowledgements}
The MeerKAT telescope is operated by the South African Radio Astronomy Observatory, which is a facility of the National Research Foundation, an agency of the Department of Science and Innovation. MeerTime data is housed and processed on the OzSTAR supercomputer at Swinburne University of Technology with the support of ADACS and the gravitational wave data centre via AAL. LSO acknowledges the support of Magdalen College, Oxford.

\bibliographystyle{mnras}
\bibliography{rvm} % if your bibtex file is called example.bib
%%%%%%%%%%%%%%%%%%%%%%%%%%%%%%%%%%%%%%%%%%%%%%%%%%

\appendix
\section{Tables}
\begin{table*}
\caption{Pulsars classified as RVM. The sign of $\beta$ is given in column 3. Column 4 indicates whether $V > L$ in more than 5 phase bins across the profile. Column 5 indicates the offset between $\phi_0$ and the profile midpoint. Interpulse pulsars are denoted with an M (main pulse) or I (interpulse) suffix.}
\label{tab:rvm}
\begin{center}
\begin{tabular}{cccccccccccccccc}
\hline
\hline
JNAME & BNAME & $\beta$ & $V_{\rm h}$ & $\Delta\phi$ \\
\hline
J0045--7042 &  & +ve & no & +ve & J0134--2937 &  & --ve & no & --ve & J0151--0635 & B0148--06 & --ve & no & 0 \\
J0304+1932 & B0301+19 & +ve & no & 0 & J0343--3000 &  & --ve & no & +ve & J0452--1759 & B0450--18 & --ve & yes & +ve \\
J0459--0210 &  & +ve & yes & --ve & J0514--4407M &  & +ve & no & +ve & J0534--6703 &  & --ve & no & +ve \\
J0536--7543 & B0538--75 & +ve & no & 0 & J0543+2329 & B0540+23 & +ve & no & +ve & J0555--7056 &  & --ve & no & +ve \\
J0614+2229 & B0611+22 & --ve & no & +ve & J0627+0649 &  & +ve & no & +ve & J0627+0706I &  & --ve & no & 0 \\
J0627+0706M &  & --ve & no & --ve & J0630--2834 & B0628--28 & +ve & no & 0 & J0631+1036 &  & +ve & no & +ve \\
J0633--2015 &  & --ve & no & 0 & J0646+0905 &  & +ve & no & --ve & J0659+1414 & B0656+14 & +ve & no & +ve \\
J0709--5923 &  & +ve & no & --ve & J0711+0931 &  & --ve & no & 0 & J0729--1448 &  & --ve & no & +ve \\
J0729--1836 & B0727--18 & --ve & no & +ve & J0737--2202 &  & --ve & no & +ve & J0742--2822 & B0740--28 & +ve & no & +ve \\
J0812--3905 &  & --ve & no & +ve & J0818--3049 &  & --ve & no & +ve & J0820--3921 &  & +ve & no & +ve \\
J0821--4221 &  & +ve & no & +ve & J0831--4406 &  & +ve & no & 0 & J0835--4510 & B0833--45 & +ve & no & +ve \\
J0837--2454 &  & --ve & yes & +ve & J0838--2621 &  & --ve & no & +ve & J0842--4851M & B0840--48M & --ve & no & --ve \\
J0843--5022 &  & --ve & yes & +ve & J0847--4316 &  & --ve & no & 0 & J0849--6322 &  & +ve & no & --ve \\
J0855--4644 &  & --ve & no & +ve & J0855--4658 &  & +ve & no & --ve & J0856--6137 & B0855--61 & --ve & no & --ve \\
J0857--4424 &  & +ve & no & --ve & J0901--4624 &  & +ve & yes & +ve & J0904--4246 & B0903--42 & +ve & no & 0 \\
J0904--7459 & B0904--74 & --ve & no & --ve & J0905--4536 &  & +ve & no & +ve & J0905--5127M &  & --ve & no & +ve \\
J0905--6019 &  & +ve & yes & 0 & J0907--5157 & B0905--51 & --ve & no & --ve & J0908--4913I & B0906--49I & --ve & no & +ve \\
J0908--4913M & B0906--49M & +ve & no & +ve & J0909--7212 & B0909--71 & +ve & no & +ve & J0912--3851 &  & +ve & yes & 0 \\
J0922+0638 & B0919+06 & --ve & no & +ve & J0924--5814 & B0923--58 & +ve & no & +ve & J0932--3217 &  & --ve & no & 0 \\
J0932--5327 &  & --ve & no & 0 & J0940--5428 &  & --ve & no & +ve & J0942--5657 & B0941--56 & --ve & no & 0 \\
J0943+1631 & B0940+16 & --ve & no & +ve & J0945--4833 &  & +ve & no & +ve & J0952--3839 & B0950--38 & --ve & no & --ve \\
J0954--5430 &  & --ve & no & 0 & J0957--5432 &  & --ve & no & +ve & J0959--4809 & B0957--47 & --ve & yes & 0 \\
J1000--5149 &  & +ve & no & --ve & J1001--5939 &  & +ve & no & 0 & J1002--5919 &  & +ve & no & +ve \\
J1012--2337 & B1010--23 & +ve & no & +ve & J1013--5934 &  & --ve & no & +ve & J1015--5719 &  & --ve & no & +ve \\
J1016--5345 & B1014--53 & --ve & no & +ve & J1016--5857 &  & +ve & no & +ve & J1020--5921 &  & --ve & no & 0 \\
J1032--5206 &  & --ve & no & 0 & J1038--5831 & B1036--58 & --ve & no & 0 & J1041--1942 & B1039--19 & +ve & no & +ve \\
J1042--5521 & B1039--55 & --ve & no & +ve & J1048--5832 & B1046--58 & --ve & no & 0 & J1049--5833 &  & --ve & no & --ve \\
J1052--6348 &  & +ve & no & --ve & J1054--6452 &  & +ve & no & --ve & J1055--6028 &  & +ve & no & 0 \\
J1055--6905 &  & +ve & no & --ve & J1057--5226M & B1055--52M & --ve & no & --ve & J1057--7914 & B1056--78 & --ve & no & 0 \\
J1103--6025 &  & --ve & no & +ve & J1104--6103 &  & --ve & no & +ve & J1105--6107 &  & --ve & no & +ve \\
J1110--5637 & B1107--56 & --ve & no & +ve & J1114--6100 & B1112--60 & +ve & no & --ve & J1115--6052 &  & +ve & no & +ve \\
J1116--4122 & B1114--41 & +ve & no & --ve & J1117--6154 &  & +ve & no & 0 & J1119--7936 & B1118--79 & --ve & no & 0 \\
J1123--6102 &  & --ve & yes & +ve & J1123--6259 &  & +ve & no & --ve & J1124--5638 &  & --ve & yes & --ve \\
J1126--6942 &  & +ve & no & --ve & J1132--4700 &  & --ve & no & +ve & J1141--3107 &  & +ve & no & --ve \\
J1141--6545 &  & --ve & yes & +ve & J1143--5536 &  & --ve & no & 0 & J1146--6030 & B1143--60 & +ve & no & +ve \\
J1148--5725 &  & +ve & no & 0 & J1151--6108 &  & --ve & no & +ve & J1152--6012 &  & +ve & no & --ve \\
J1159--6409 &  & --ve & no & 0 & J1204--6843 &  & +ve & yes & +ve & J1211--6324 &  & --ve & no & +ve \\
J1215--5328 &  & +ve & yes & +ve & J1220--6318 &  & +ve & no & +ve & J1222--5738 &  & --ve & yes & --ve \\
J1224--6208 &  & --ve & no & +ve & J1236--5033 &  & --ve & no & +ve & J1245--6238 &  & +ve & yes & 0 \\
J1252--6314 &  & +ve & no & --ve & J1253--5820 &  & --ve & no & +ve & J1254--6150 &  & +ve & no & --ve \\
J1301--6310 &  & --ve & no & +ve & J1302--6350 & B1259--63 & --ve & no & --ve & J1305--6203 &  & +ve & no & --ve \\
J1308--4650 &  & --ve & no & 0 & J1311--1228 & B1309--12 & +ve & no & --ve & J1312--5402 & B1309--53 & --ve & no & 0 \\
J1312--5516 & B1309--55 & --ve & no & 0 & J1319--6105 &  & --ve & no & --ve & J1320--5359 & B1317--53 & +ve & no & +ve \\
J1326--6700 & B1322--66 & --ve & no & 0 & J1327--6301 & B1323--627 & +ve & no & +ve & J1328--4357 & B1325--43 & --ve & no & 0 \\
J1331--5245 &  & +ve & yes & 0 & J1339--6618 &  & --ve & no & +ve & J1340--6456 & B1336--64 & --ve & yes & +ve \\
J1345--6115 &  & +ve & yes & 0 & J1350--5115 &  & +ve & no & --ve & J1352--6803 &  & +ve & no & --ve \\
J1357--6429 &  & --ve & no & --ve & J1401--6357 & B1358--63 & --ve & no & 0 & J1403--6310 &  & --ve & no & +ve \\
J1403--7646 &  & +ve & no & +ve & J1404+1159 &  & +ve & no & 0 & J1412--6111 &  & +ve & no & 0 \\
J1413--6307I & B1409--62I & +ve & no & --ve & J1414--6802 &  & --ve & no & 0 & J1415--6621 &  & +ve & no & +ve \\
J1416--6037 &  & --ve & no & +ve & J1420--6048 &  & --ve & no & +ve & J1424--5556 &  & --ve & no & --ve \\
J1424--6438 &  & +ve & no & 0 & J1425--6210 &  & --ve & no & --ve & J1427--4158 &  & --ve & no & 0 \\
J1428--5530 & B1424--55 & --ve & yes & 0 & J1432--5032 &  & --ve & no & --ve & J1435--5954 &  & --ve & no & +ve \\
J1443--5122 &  & --ve & no & 0 & J1452--5851 &  & +ve & no & +ve & J1502--5653 &  & +ve & no & +ve \\
J1502--6128 &  & +ve & no & +ve & J1507--4352 & B1504--43 & +ve & no & +ve & J1512--5431 &  & +ve & no & 0 \\
J1513--5739 &  & --ve & no & 0 & J1517--4636 &  & --ve & no & +ve & J1518--0627 &  & --ve & no & --ve \\
\hline
\end{tabular}
\end{center}
\end{table*}
\begin{table*}
\addtocounter{table}{-1}
\caption{Pulsars classified as RVM (continued).}
\begin{center}
\begin{tabular}{cccccccccccccccc}
\hline
\hline
JNAME & BNAME & $\beta$ & $V_{\rm h}$ & $\Delta\phi$ \\
\hline
J1519--6106 &  & --ve & no & +ve & J1519--6308 &  & --ve & no & +ve & J1522--5525 &  & +ve & no & 0 \\
J1524--5819 &  & --ve & no & --ve & J1527--3931 & B1524--39 & +ve & yes & 0 & J1530--5327 &  & --ve & no & 0 \\
J1531--4012 &  & +ve & no & 0 & J1535--4114 &  & --ve & no & +ve & J1535--4415 &  & --ve & no & --ve \\
J1535--5848 &  & +ve & no & +ve & J1537--4912 &  & +ve & no & +ve & J1537--5153 &  & --ve & no & 0 \\
J1538+2345 &  & --ve & no & 0 & J1538--5732 &  & --ve & no & +ve & J1539--4828 &  & +ve & no & 0 \\
J1539--6322 &  & +ve & no & 0 & J1542--5303 &  & +ve & no & +ve & J1543--5013 &  & --ve & no & 0 \\
J1548--4927 &  & --ve & no & +ve & J1549+2113 &  & --ve & no & --ve & J1549--4848I &  & --ve & no & +ve \\
J1550--5242 &  & --ve & no & +ve & J1554--5209 &  & +ve & no & 0 & J1555--2341 & B1552--23 & --ve & no & +ve \\
J1558--5756 &  & +ve & no & 0 & J1559--4438 & B1556--44 & +ve & yes & --ve & J1601--5335 &  & --ve & no & +ve \\
J1603--5657 &  & --ve & no & --ve & J1605--5257 & B1601--52 & +ve & no & +ve & J1611--4949 &  & --ve & no & --ve \\
J1614+0737 & B1612+07 & +ve & no & --ve & J1614--3937 &  & +ve & yes & 0 & J1615--5444 &  & --ve & no & +ve \\
J1616--5017 &  & --ve & yes & 0 & J1617--4216 &  & --ve & no & 0 & J1622--3751 &  & +ve & no & 0 \\
J1622--4347 &  & --ve & no & +ve & J1622--4802 &  & +ve & no & --ve & J1622--4950 &  & --ve & yes & --ve \\
J1623--4949 &  & --ve & no & --ve & J1626--4537 &  & --ve & no & --ve & J1632--1013 &  & +ve & no & +ve \\
J1634--5640 &  & +ve & no & 0 & J1635--4944 &  & +ve & no & --ve & J1636--2614 &  & --ve & no & +ve \\
J1637--4642 &  & +ve & no & +ve & J1638--3951 &  & +ve & no & --ve & J1638--4344 &  & +ve & no & 0 \\
J1646--5123 &  & --ve & no & +ve & J1648--6044 &  & --ve & yes & --ve & J1653--3838 & B1650--38 & --ve & no & --ve \\
J1655--3844 &  & --ve & yes & 0 & J1656--3621 &  & --ve & no & +ve & J1700--3312 &  & --ve & no & 0 \\
J1700--3919 &  & +ve & yes & 0 & J1703--1846 & B1700--18 & +ve & no & +ve & J1703--3241 & B1700--32 & --ve & no & 0 \\
J1703--4442 &  & --ve & no & 0 & J1704--5236 &  & +ve & no & --ve & J1705--3950 &  & +ve & yes & +ve \\
J1707--4341 &  & +ve & yes & +ve & J1707--4417 &  & --ve & no & 0 & J1709--1640 & B1706--16 & +ve & no & 0 \\
J1709--3626 &  & --ve & no & 0 & J1709--4401 &  & --ve & no & 0 & J1711--1509 & B1709--15 & --ve & yes & +ve \\
J1716--4711 &  & +ve & yes & 0 & J1718--3825 &  & --ve & no & +ve & J1719--4302 &  & +ve & no & +ve \\
J1720--0212 & B1718--02 & +ve & yes & +ve & J1720--1633 & B1717--16 & --ve & no & 0 & J1720--2933 & B1717--29 & +ve & yes & 0 \\
J1722--3207 & B1718--32 & +ve & no & --ve & J1722--3712I & B1719--37I & --ve & no & +ve & J1722--3712M & B1719--37M & --ve & no & +ve \\
J1722--4400 &  & +ve & no & 0 & J1723--3659 &  & +ve & yes & +ve & J1727--2739 &  & --ve & no & +ve \\
J1728--0007 & B1726--00 & --ve & no & --ve & J1733--3716 & B1730--37 & +ve & yes & +ve & J1733--4005 &  & --ve & yes & --ve \\
J1734--0212 & B1732--02 & --ve & yes & 0 & J1737--3102 &  & +ve & yes & --ve & J1737--3555 & B1734--35 & +ve & no & 0 \\
J1738--2955 &  & +ve & no & --ve & J1739--2903I & B1736--29I & +ve & no & +ve & J1739--3023 &  & --ve & no & +ve \\
J1740+1000 &  & --ve & no & +ve & J1740+1311 & B1737+13 & +ve & no & +ve & J1740--3015 & B1737--30 & +ve & yes & +ve \\
J1741--0840 & B1738--08 & --ve & no & 0 & J1743--0339 & B1740--03 & --ve & no & +ve & J1743--3150 & B1740--31 & --ve & no & 0 \\
J1743--4212 &  & --ve & no & +ve & J1744--3130 &  & +ve & no & +ve & J1745--3812 &  & --ve & no & 0 \\
J1746+2245 &  & --ve & yes & --ve & J1748--1300 & B1745--12 & +ve & yes & +ve & J1750--3503 &  & +ve & no & --ve \\
J1751--3323 &  & +ve & no & --ve & J1754--3443 &  & +ve & yes & --ve & J1757--1500 &  & --ve & yes & +ve \\
J1759--3107 &  & +ve & no & +ve & J1800--0125 &  & +ve & no & +ve & J1801--2920 & B1758--29 & --ve & yes & 0 \\
J1802+0128 &  & +ve & no & --ve & J1803--2137 & B1800--21 & +ve & no & +ve & J1806--1154 & B1804--12 & +ve & yes & 0 \\
J1808--0813 &  & +ve & no & --ve & J1808--3249 &  & +ve & no & +ve & J1809--1429 &  & --ve & yes & --ve \\
J1809--1917 &  & +ve & no & --ve & J1809--1943 &  & +ve & yes & +ve & J1809--2109 & B1806--21 & --ve & no & 0 \\
J1810--5338 & B1806--53 & --ve & no & 0 & J1813--2113 &  & --ve & no & --ve & J1816--0755M &  & --ve & no & 0 \\
J1817--3618 & B1813--36 & --ve & yes & +ve & J1819+1305 &  & --ve & no & +ve & J1819--0925 &  & +ve & no & 0 \\
J1819--1008 &  & +ve & no & --ve & J1819--1458 &  & --ve & no & 0 & J1821--0331 &  & --ve & no & --ve \\
J1823--0154 &  & +ve & no & +ve & J1823--3106 & B1820--31 & +ve & no & +ve & J1824--0127 &  & --ve & no & +ve \\
J1826--1334 & B1823--13 & +ve & yes & +ve & J1828+1359 &  & +ve & yes & --ve & J1830--1059 & B1828--11 & +ve & no & +ve \\
J1831--0823 &  & --ve & yes & 0 & J1831--0952 &  & --ve & no & +ve & J1835--0349 &  & --ve & yes & +ve \\
J1835--0944 &  & +ve & no & --ve & J1835--1020 &  & +ve & no & 0 & J1835--1106 &  & --ve & no & +ve \\
J1837+1221 &  & +ve & no & 0 & J1837--0045 &  & +ve & no & --ve & J1837--1837 &  & +ve & no & +ve \\
J1839--0436 &  & +ve & yes & +ve & J1839--1238 &  & +ve & yes & --ve & J1840+0214 &  & --ve & no & --ve \\
J1840--0809 &  & --ve & no & +ve & J1840--1122 &  & --ve & no & +ve & J1841+0912 & B1839+09 & +ve & no & 0 \\
J1841--0345 &  & --ve & no & +ve & J1841--0524 &  & +ve & no & +ve & J1842+0257 &  & +ve & no & 0 \\
J1842+0358 &  & --ve & no & --ve & J1842+0638 &  & +ve & no & --ve & J1842--0359 & B1839--04 & --ve & no & 0 \\
J1842--0905 &  & --ve & yes & +ve & J1843--0000 &  & +ve & no & +ve & J1843--0211 &  & --ve & no & 0 \\
J1843--0702I &  & +ve & no & --ve & J1843--0806 &  & +ve & no & --ve & J1845--0434 & B1842--04 & --ve & yes & 0 \\
J1845--0545 &  & +ve & no & --ve & J1845--0635 &  & +ve & no & --ve & J1845--1114 &  & --ve & no & +ve \\
J1846+0051 &  & +ve & no & +ve & J1846--07492 &  & +ve & yes & 0 & J1847--0438 &  & --ve & no & --ve \\
J1847--0605 &  & +ve & no & +ve & J1848+0604 &  & --ve & yes & +ve & J1849+0409M &  & --ve & no & +ve \\
\hline
\end{tabular}
\end{center}
\end{table*}
\begin{table*}
\addtocounter{table}{-1}
\caption{Pulsars classified as RVM (continued).}
\begin{center}
\begin{tabular}{cccccccccccccccc}
\hline
\hline
JNAME & BNAME & $\beta$ & $V_{\rm h}$ & $\Delta\phi$ \\
\hline
J1849+2423 &  & +ve & no & +ve & J1849--0317 &  & --ve & no & 0 & J1849--0614 &  & +ve & no & +ve \\
J1850+0026 &  & +ve & yes & --ve & J1850+1335 & B1848+13 & +ve & no & 0 & J1851+1259 & B1848+12 & --ve & no & 0 \\
J1851--0053 &  & +ve & yes & 0 & J1852--0635 &  & +ve & no & 0 & J1854+0319 &  & +ve & yes & --ve \\
J1854--1421 & B1851--14 & +ve & no & --ve & J1855+0307 &  & --ve & no & 0 & J1855--0941 &  & --ve & no & +ve \\
J1856--0526 &  & --ve & yes & +ve & J1857+0057 & B1854+00 & --ve & no & +ve & J1900--0051 &  & --ve & no & +ve \\
J1900--0933 &  & +ve & no & --ve & J1901+0124 &  & --ve & no & +ve & J1901+0234 &  & +ve & no & +ve \\
J1901+0716 & B1859+07 & --ve & no & --ve & J1901+1306 &  & --ve & no & 0 & J1901--0312 &  & +ve & no & +ve \\
J1902+0615 & B1900+06 & --ve & no & --ve & J1903+2225 &  & --ve & no & 0 & J1903--0632 & B1900--06 & +ve & no & +ve \\
J1903--0848 &  & +ve & no & +ve & J1904+1011 & B1901+10 & +ve & no & +ve & J1905+0616 &  & +ve & no & 0 \\
J1905+0709 & B1903+07 & --ve & no & +ve & J1907+0631 &  & +ve & no & +ve & J1907+0740 &  & --ve & no & +ve \\
J1907+0918 &  & --ve & yes & 0 & J1907+1149 &  & --ve & no & +ve & J1908+0500 &  & +ve & no & +ve \\
J1909+0749I &  & +ve & no & --ve & J1909+0749M &  & --ve & no & +ve & J1910+0728 &  & --ve & yes & 0 \\
J1911+1758 &  & +ve & no & --ve & J1912+2104 & B1910+20 & --ve & no & 0 & J1913+0936 & B1911+09 & --ve & no & +ve \\
J1914+0219 &  & +ve & no & 0 & J1914+1122 & B1911+11 & +ve & no & 0 & J1915+0738 &  & --ve & no & +ve \\
J1915+0752 &  & +ve & no & +ve & J1915+1606 & B1913+16 & --ve & no & +ve & J1915+1647 & B1913+167 & +ve & no & 0 \\
J1916+1312 & B1914+13 & --ve & yes & +ve & J1917+0834 &  & --ve & no & 0 & J1917+1353 & B1915+13 & +ve & no & +ve \\
J1918+1444 & B1916+14 & --ve & no & 0 & J1918+1541M &  & --ve & no & --ve & J1918--1052 &  & --ve & yes & +ve \\
J1919+1745 &  & --ve & no & 0 & J1920--0950 &  & --ve & no & +ve & J1921+0812 &  & --ve & no & +ve \\
J1921+1419 & B1919+14 & +ve & no & +ve & J1922+1733 &  & +ve & no & +ve & J1924+1631 &  & --ve & no & --ve \\
J1925+1720 &  & +ve & no & +ve & J1926+1434 & B1924+14 & --ve & yes & +ve & J1926+1648 & B1924+16 & --ve & no & +ve \\
J1926+2016 &  & +ve & no & +ve & J1926--0652 &  & +ve & no & 0 & J1927+1852 & B1925+18 & +ve & no & +ve \\
J1927+2234 & B1925+22 & --ve & no & 0 & J1928+1746 &  & --ve & no & +ve & J1931+1439 &  & +ve & no & 0 \\
J1931+1536 & B1929+15 & --ve & no & +ve & J1932+1059 & B1929+10 & +ve & no & --ve & J1932+1500 &  & +ve & no & 0 \\
J1932+2220 & B1930+22 & --ve & no & +ve & J1934+2352 &  & --ve & no & +ve & J1935+2025I &  & --ve & no & --ve \\
J1935+2025M &  & --ve & no & +ve & J1938+0650 &  & --ve & no & +ve & J1938+2010 &  & --ve & no & +ve \\
J1940+2245 &  & +ve & no & +ve & J1940+2337 &  & +ve & no & +ve & J1941+0121 &  & +ve & no & --ve \\
J1943--1237 & B1940--12 & +ve & no & +ve & J1944+1755 & B1942+17 & --ve & no & --ve & J1946--1312 &  & --ve & no & 0 \\
J1949--2524 & B1946--25 & --ve & no & +ve & J1951+1123 &  & +ve & no & +ve & J1956+0838 &  & +ve & no & --ve \\
J2002+1637 &  & +ve & no & +ve & J2005--0020 &  & +ve & no & 0 & J2007+0910 &  & --ve & no & +ve \\
J2017+2043 &  & +ve & yes & --ve & J2038--3816 &  & --ve & no & 0 & J2043+2740 &  & +ve & no & +ve \\
J2048--1616 & B2045--16 & +ve & no & 0 & J2108--3429 &  & +ve & yes & 0 & J2111+2106 &  & +ve & yes & +ve \\
J2124+1407 & B2122+13 & --ve & no & 0 & J2127--6648 & B2123--67 & --ve & no & 0 & J2139+2242 &  & --ve & no & 0 \\
J2155--3118 & B2152--31 & +ve & yes & 0 & J2215+1538 &  & --ve & no & --ve & J2234+2114 &  & --ve & yes & --ve \\
J2248--0101 &  & +ve & no & 0 & J2324--6054 & B2321--61 & +ve & no & 0 & J2346--0609 &  & +ve & no & 0 \\
 \\
\hline
\end{tabular}
\end{center}
\end{table*}
\begin{table*}
\caption{Pulsars classified as flat. The sign of $\beta$ is given in column 3. Column 4 indicates whether $V > L$ in more than 5 phase bins across the profile. Interpulse pulsars are denoted with an M (main pulse) or I (interpulse) suffix.}
\label{tab:flat}
\begin{center}
\begin{tabular}{cccccccccccc}
\hline
\hline
JNAME & BNAME & $\beta$ & $V_{\rm h}$\\
\hline
J0108--1431 &  & --ve & no & J0448--2749 &  & +ve & yes & J0631+0646 &  & --ve & no \\
J0934--4154 &  & +ve & no & J1016--5819 &  & --ve & no & J1047--3032 &  & +ve & no \\
J1047--6709 &  & +ve & no & J1052--5954 &  & +ve & no & J1054--5943 &  & --ve & no \\
J1056--6258 & B1054--62 & +ve & no & J1119--6127 &  & +ve & no & J1137--6700 &  & --ve & no \\
J1154--6250 &  & +ve & no & J1210--6550 &  & +ve & no & J1232--4742 &  & --ve & no \\
J1249--6507 &  & --ve & no & J1301--6305 &  & --ve & no & J1339--4712 &  & --ve & yes \\
J1347--5947 &  & +ve & no & J1355--6206 &  & +ve & no & J1406--5806 &  & --ve & no \\
J1424--5822 &  & +ve & yes & J1452--6036 &  & --ve & no & J1509--6015 &  & --ve & no \\
J1511--5414 &  & --ve & no & J1513--5908 & B1509--58 & --ve & no & J1524--5625 &  & --ve & no \\
J1531--5610 &  & +ve & no & J1551--4424 &  & +ve & no & J1603--2531 &  & +ve & no \\
J1627--5547 &  & --ve & no & J1638--4417 &  & +ve & no & J1643--4505 &  & +ve & yes \\
J1649--4349 &  & --ve & no & J1650--4921 &  & --ve & no & J1652--1400 &  & --ve & no \\
J1658--4958 &  & +ve & no & J1701--3130 &  & --ve & no & J1709--4429 & B1706--44 & --ve & no \\
J1715--3903 &  & +ve & no & J1718--4539 &  & --ve & no & J1739--3951 &  & --ve & no \\
J1740--3052 &  & --ve & no & J1742--4616 &  & --ve & no & J1811+0702 &  & --ve & no \\
J1813+1822 &  & --ve & no & J1827--0934 &  & --ve & no & J1832--0644 &  & --ve & no \\
J1833--1034 &  & --ve & no & J1836--1324 &  & +ve & no & J1839--0402 &  & --ve & no \\
J1842+1332 &  & --ve & no & J1843--0510 &  & +ve & no & J1851+0418 & B1848+04 & --ve & no \\
J1856+0113 & B1853+01 & --ve & no & J1901+0355 &  & +ve & no & J1904+0004 &  & --ve & yes \\
J1904+0800 &  & --ve & no & J1904--0150 &  & +ve & no & J1906+0649 &  & --ve & no \\
J1907+0345 &  & --ve & no & J1912+1036 & B1910+10 & +ve & no & J1913+0832M &  & --ve & no \\
J1913+0904 &  & --ve & no & J1913+1000 &  & --ve & no & J1924+1639 &  & +ve & no \\
J1928+1443 &  & +ve & no & J1929+2121 &  & +ve & no & J1935+1159 &  & --ve & no \\
J1946+1805 & B1944+17 & --ve & no & J2307+2225 &  & +ve & yes &  \\
\hline
\end{tabular}
\end{center}
\end{table*}
\begin{table*}
\caption{Pulsars classified as non--RVM. Interpulse pulsars are denoted with an M (main pulse) or I (interpulse) suffix.}
\label{tab:bad}
\begin{center}
\begin{tabular}{cc|cc|cc|cc|cc}
\hline
\hline
JNAME & BNAME \\
\hline
J0034--0721 & B0031--07 & J0152--1637 & B0149--16 & J0206--4028 & B0203--40 & J0211--8159 &  & J0255--5304 & B0254--53 \\
J0302+2252 &  & J0401--7608 & B0403--76 & J0421--0345 &  & J0450--1248 & B0447--12 & J0517+2212 &  \\
J0520--2553 &  & J0525+1115 & B0523+11 & J0529--6652 & B0529--66 & J0533+0402 &  & J0540--7125 &  \\
J0601--0527 & B0559--05 & J0621+0336 &  & J0624--0424 & B0621--04 & J0629+2415 & B0626+24 & J0630--0046 &  \\
J0647+0913 &  & J0656--2228 &  & J0719--2545 &  & J0725--1635 &  & J0726--2612 &  \\
J0733--2345 &  & J0738--4042 & B0736--40 & J0746--4529 &  & J0749--4247 &  & J0758--1528 & B0756--15 \\
J0807--5421 &  & J0809--4753 & B0808--47 & J0818--3232 &  & J0820--1350 & B0818--13 & J0823+0159 & B0820+02 \\
J0835--3707 &  & J0837+0610 & B0834+06 & J0837--4135 & B0835--41 & J0840--5332 & B0839--53 & J0846--3533 & B0844--35 \\
J0855--3331 & B0853--33 & J0902--6325 & B0901--63 & J0905--5127I &  & J0908--1739 & B0906--17 & J0919--6040 &  \\
J0922--4949 &  & J0924--5302 & B0922--52 & J0934--5249 & B0932--52 & J0941--5244 &  & J0942--5552 & B0940--55 \\
J0944--1354 & B0942--13 & J0949--6902 &  & J0953+0755 & B0950+08 & J0955--5304 & B0953--52 & J1001--5507 & B0959--54 \\
J1001--5559 &  & J1003--4747 & B1001--47 & J1012--5857 & B1011--58 & J1017--5621 & B1015--56 & J1018--1642 & B1016--16 \\
J1034--3224 &  & J1035--6345 &  & J1036--4926 &  & J1036--6559 &  & J1043--6116 &  \\
J1046--5813 & B1044--57 & J1057--5226I & B1055--52I & J1058--5957 &  & J1059--5742 & B1056--57 & J1112--6613 & B1110--65 \\
J1112--6926 & B1110--69 & J1121--5444 & B1119--54 & J1123--4844 &  & J1123--6651 &  & J1126--6054M & B1124--60M \\
J1132--5627 &  & J1133--6250 & B1131--62 & J1136+1551 & B1133+16 & J1136--5525 & B1133--55 & J1141--3322 &  \\
J1142--6230 &  & J1144--6217 &  & J1156--5707 &  & J1157--6224 & B1154--62 & J1159--7910 &  \\
J1202--5820 & B1159--58 & J1210--5559 &  & J1224--6407 & B1221--63 & J1225--5556 &  & J1225--6035 &  \\
J1225--6408 & B1222--63 & J1231--4609 &  & J1235--6354 &  & J1237--6725 &  & J1239+2453 & B1237+25 \\
J1239--6832 & B1236--68 & J1240--4124 & B1237--41 & J1243--6423 & B1240--64 & J1244--5053 &  & J1246+2253 &  \\
J1248--6444 &  & J1251--7407 &  & J1257--1027 & B1254--10 & J1259--6741 & B1256--67 & J1303--6305 &  \\
J1305--6455 & B1302--64 & J1306--6617 & B1303--66 & J1308--5844 &  & J1312--6400 &  & J1313+0931 &  \\
J1319--6056 & B1316--60 & J1322--6241 &  & J1324--6302 &  & J1326--5859 & B1323--58 & J1326--6408 & B1323--63 \\
J1327--6222 & B1323--62 & J1328--4921 & B1325--49 & J1334--5839 &  & J1336--2522 &  & J1338--6204 & B1334--61 \\
J1341--6023 &  & J1355--5153 & B1352--51 & J1355--5925 &  & J1356--5521 &  & J1357--62 & B1353--62 \\
J1405--5641 &  & J1410--7404 &  & J1413--6307M & B1409--62M & J1418--3921 &  & J1420--5416 & B1417--54 \\
J1423--6953 &  & J1429--5935 &  & J1430--6623 & B1426--66 & J1440--6344 & B1436--63 & J1449--5846 &  \\
J1453--6413 & B1449--64 & J1456--6843 & B1451--68 & J1501--0046 &  & J1507--6640 & B1503--66 & J1514--4834 & B1510--48 \\
J1517--4356 &  & J1518--3952 &  & J1522--5829 & B1518--58 & J1524--5706 &  & J1525--5417 &  \\
J1527--5552 & B1523--55 & J1528--4109 &  & J1530--6343 &  & J1534--4428 &  & J1534--5334 & B1530--53 \\
J1534--5405 & B1530--539 & J1536--3602 &  & J1539--5626 & B1535--56 & J1542--5034 &  & J1543+0929 & B1541+09 \\
J1543--0620 & B1540--06 & J1544--5308 & B1541--52 & J1547--5750 &  & J1551--6214 &  & J1555--3134 & B1552--31 \\
J1557--4258 &  & J1600--5751 & B1556--57 & J1602--5100 & B1558--50 & J1603--2712 & B1600--27 & J1604--4909 & B1600--49 \\
J1604--7203 &  & J1607--0032 & B1604--00 & J1609--4616 &  & J1611--5209M & B1607--52M & J1612+2008 &  \\
J1612--5805 &  & J1613--4714 & B1609--47 & J1615--5537 & B1611--55 & J1617--4608 &  & J1618--4723 &  \\
J1621--5039 &  & J1622--4332 &  & J1623--0908 & B1620--09 & J1624--4411 &  & J1625--4048 &  \\
J1626--6621 &  & J1627--5936 &  & J1632--4621 &  & J1635--5954 & B1630--59 & J1637--4553 & B1634--45 \\
J1638--3815 &  & J1638--5226 &  & J1639--4604 & B1635--45 & J1645--0317 & B1642--03 & J1646--6831 & B1641--68 \\
J1648--3256 &  & J1649--3805 &  & J1649--3935 &  & J1650--1654 &  & J1651--1709 & B1648--17 \\
J1651--4246 & B1648--42 & J1651--5222 & B1647--52 & J1654--3710 &  & J1655--3048 &  & J1659--1305 & B1657--13 \\
J1700--4939 &  & J1703--4851 &  & J1705--1906I & B1702--19I & J1705--1906M & B1702--19M & J1705--3423 &  \\
J1707--4729 &  & J1711--5350 & B1707--53 & J1714--1054 &  & J1717--5800 &  & J1720+2150 &  \\
J1731--4744 & B1727--47 & J1735--0724 & B1732--07 & J1738--3211 & B1735--32 & J1739+0612 &  & J1739--1313 &  \\
J1739--2903M & B1736--29M & J1741--3927 & B1737--39 & J1743--1351 & B1740--13 & J1744--1610 &  & J1745--0129 &  \\
J1745--3040 & B1742--30 & J1750--3157 & B1747--31 & J1751--4657 & B1747--46 & J1752--2806 & B1749--28 & J1754--3510 &  \\
J1755--0903M &  & J1757--2421 & B1754--24 & J1801--0357 & B1758--03 & J1803--3329 &  & J1805+0306 & B1802+03 \\
J1805--0619 &  & J1806+1023 &  & J1807--0847 & B1804--08 & J1807--2715 & B1804--27 & J1808--1020 &  \\
J1809--0119 &  & J1809--0743 &  & J1811--0154 &  & J1811--4930 &  & J1812--1718 & B1809--173 \\
J1812--3039 &  & J1814--0521 &  & J1817--3837 &  & J1818--0151 &  & J1820--0427 & B1818--04 \\
J1820--0509 &  & J1820--1818 & B1817--18 & J1821+1715 &  & J1822+0705 &  & J1822--4209 &  \\
J1823+0550 & B1821+05 & J1824--1945 & B1821--19 & J1825+0004 & B1822+00 & J1825--0935 & B1822--09 & J1826--1131 & B1823--11 \\
J1827--0750 &  & J1828--0611 &  & J1829--1751 & B1826--17 & J1832--0827 & B1829--08 & J1833--0338 & B1831--03 \\
J1833--0827 & B1830--08 & J1834--0031 &  & J1834--1202 &  & J1836--0436 & B1834--04 & J1836--1008 & B1834--10 \\
J1837--0653 & B1834--06 & J1838+1523 &  & J1838+1650 &  & J1839--0905 &  & J1840--0815 &  \\
J1841--0157 &  & J1841--0425 & B1838--04 & J1843--0459 &  & J1843--1507 &  & J1844+1454 & B1842+14 \\
\hline
\end{tabular}
\end{center}
\end{table*}
\begin{table*}
\addtocounter{table}{-1}
\caption{Pulsars classified as non--RVM (continued).}
\begin{center}
\begin{tabular}{cc|cc|cc|cc|cc}
\hline
\hline
JNAME & BNAME \\
\hline
J1844--0433 & B1841--04 & J1845+0623 &  & J1845--0743 &  & J1847--0402 & B1844--04 & J1848+1516 &  \\
J1848--0023 &  & J1848--0123 & B1845--01 & J1848--1150 &  & J1849--0636 & B1846--06 & J1852--0118 &  \\
J1852--2610 &  & J1853--0004 &  & J1857+0212 & B1855+02 & J1859+1526 &  & J1900--2600 & B1857--26 \\
J1900--7951 & B1851--79 & J1901+0156 & B1859+01 & J1901+0331 & B1859+03 & J1901--0906 &  & J1902+0556 & B1900+05 \\
J1903+0135 & B1900+01 & J1904--1224 &  & J1905--0056 & B1902--01 & J1906+0641 & B1904+06 & J1907--1532 &  \\
J1909+0007 & B1907+00 & J1909+0254 & B1907+02 & J1909+1102 & B1907+10 & J1909+1859 &  & J1910+0358 & B1907+03 \\
J1910+0714 &  & J1910+1231 & B1907+12 & J1910--0309 & B1907--03 & J1913+1400 & B1911+13 & J1913--0440 & B1911--04 \\
J1915+0227 &  & J1915+1009 & B1913+10 & J1916+0951 & B1914+09 & J1916--2939 &  & J1919+0021 & B1917+00 \\
J1919+0134 &  & J1921+1948 & B1918+19 & J1921+2003 & B1919+20 & J1921+2153 & B1919+21 & J1923+1706 & B1921+17 \\
J1926+0431 & B1923+04 & J1929+1955 &  & J1930+1316 & B1927+13 & J1930--1852 &  & J1932--3655 &  \\
J1933+1304 & B1930+13 & J1935+1616 & B1933+16 & J1936+1536 & B1933+15 & J1938+2213 &  & J1941+1341 &  \\
J1941--2602 & B1937--26 & J1943+0609 &  & J1945--0040 & B1942--00 & J1946+2244 & B1944+22 & J1946--2913 & B1943--29 \\
J1947+0915 &  & J1949+2306 &  & J1952+1410 & B1949+14 & J2006--0807 & B2003--08 & J2037+1942 & B2034+19 \\
J2045+0912 &  & J2046+1540 & B2044+15 & J2046--0421 & B2043--04 & J2053--7200 & B2048--72 & J2116+1414 & B2113+14 \\
J2136--1606 &  & J2144--3933 &  & J2154--2812 &  & J2253+1516 &  & J2317+2149 & B2315+21 \\
J2330--2005 & B2327--20 &  \\
\hline
\end{tabular}
\end{center}
\end{table*}
\begin{table*}
\caption{Pulsars for which an RVM fit was not attempted. Interpulse pulsars are denoted with an M (main pulse) or I (interpulse) suffix.}
\label{tab:no}
\begin{center}
\begin{tabular}{cccccccccc}
\hline
\hline
PSR \\
\hline
J0038--2501 &  & J0045--7319 &  & J0111--7131 &  & J0113--7220 &  & J0131--7310 &  \\
J0133--6957 &  & J0137+1654 &  & J0418--4154 &  & J0449--7031 &  & J0455--6951 & B0456--69 \\
J0456--7031 &  & J0457--6337 &  & J0502--6617 & B0502--66 & J0511--6508 &  & J0514--4407I &  \\
J0519--6932 &  & J0522--6847 &  & J0532--6639 &  & J0540--6919 & B0540--69 & J0543--6851 &  \\
J0623+0340 &  & J0628+0909 &  & J0633+1746 &  & J0636--4549 &  & J0652--0142 &  \\
J0656--5449 &  & J0658+0022 &  & J0804--3647 &  & J0808--3937 &  & J0815+0939 &  \\
J0820--3826 &  & J0820--4114 & B0818--41 & J0828--3417 & B0826--34 & J0834--4159 &  & J0836--4233 &  \\
J0927+2345 &  & J0930--2301 &  & J0943+2253 &  & J1006--6311 &  & J1012--5830 &  \\
J1019--5749 &  & J1020--6026 &  & J1021--5601 &  & J1028--5819 &  & J1031--6117 &  \\
J1048--5838 &  & J1054--5946 &  & J1055--6236 &  & J1056--5709 &  & J1057--4754 &  \\
J1106--6438 &  & J1107--5907 &  & J1112--6103 &  & J1116--2444 &  & J1124--5916 &  \\
J1124--6421 &  & J1126--6054I & B1124--60I & J1130--5826 &  & J1130--5925 &  & J1138--6207 &  \\
J1143--5158 &  & J1156--5909 &  & J1201--6306 &  & J1214--5830 &  & J1216--6223 &  \\
J1233--6312 &  & J1233--6344 &  & J1234--3630 &  & J1243--5735 &  & J1244--6359 &  \\
J1248--6344 &  & J1255--6131 &  & J1306--6242 &  & J1309--6526 &  & J1316--6232 &  \\
J1317--5759 &  & J1317--6302 &  & J1320--3512 &  & J1321--5922 &  & J1327--6400 &  \\
J1332--3032 &  & J1333--4449 &  & J1337--6306 &  & J1341--6220 & B1338--62 & J1346--4918 &  \\
J1349--6130 &  & J1359--6038 & B1356--60 & J1400--6325 &  & J1406--6121 &  & J1407--6048 &  \\
J1409--6953 &  & J1410--6132 &  & J1412--6145 &  & J1413--6141 &  & J1413--6222 &  \\
J1425--5723 &  & J1425--5759 &  & J1433--6038 &  & J1434--6006 &  & J1437--6146 &  \\
J1444--6026 &  & J1457--5902 &  & J1503+2111 &  & J1504--5621 &  & J1509--5850 &  \\
J1511--5835 &  & J1512--5759 & B1508--57 & J1513--5946 &  & J1514--5925 &  & J1515--5720 &  \\
J1518--5415 &  & J1519--5734 &  & J1525--5523 &  & J1525--5605 &  & J1535--5450 &  \\
J1536--5907 &  & J1538--5438 &  & J1538--5551 &  & J1538--5621 &  & J1538--5750 &  \\
J1539--5521 &  & J1543--5459 &  & J1546--5302 &  & J1547--5839 &  & J1548--4821 &  \\
J1548--5607 &  & J1549--4848M &  & J1549--5722 &  & J1550--5418 &  & J1551--5310 &  \\
J1555--0515 &  & J1600--5044 & B1557--50 & J1603--3539 &  & J1604--4718 &  & J1607--6449 &  \\
J1610--5006 &  & J1611--5209I & B1607--52I & J1611--5847 &  & J1613--5211 &  & J1614--5048 & B1610--50 \\
J1617--5055 &  & J1620--5414 &  & J1621--5243 &  & J1623--0841 &  & J1625--4904 &  \\
J1625--4913 &  & J1627+1419 &  & J1627--4706 &  & J1630--4719 &  & J1630--4733 & B1626--47 \\
J1632--4757 &  & J1632--4818 &  & J1633--4453 & B1630--44 & J1633--5015 & B1629--50 & J1634--5107 &  \\
J1635+2332 &  & J1636--4440 &  & J1637--4450 &  & J1638--4608 &  & J1638--4725 &  \\
J1640--4648 &  & J1640--4715 & B1636--47 & J1640--4951 &  & J1643--4550 &  & J1644--4657 &  \\
J1645+1012 &  & J1647--3607 &  & J1648--4611 &  & J1649--4653 &  & J1650--4126 &  \\
J1650--4341 &  & J1650--4502 &  & J1653--4249 &  & J1657--4432 &  & J1659--4439 &  \\
J1700--4012 &  & J1700--4422 &  & J1701--4533 & B1657--45 & J1702--4128 &  & J1702--4306 &  \\
J1705--4331 &  & J1706--6118 &  & J1707--4053 & B1703--40 & J1708--3641 &  & J1709--3841 &  \\
J1710--4148 &  & J1711--4322 &  & J1715--3859 &  & J1716--4111 &  & J1717--3425 & B1714--34 \\
J1717--3737 &  & J1717--4054 & B1713--40 & J1719--4006 & B1715--40 & J1720--3659 &  & J1721--3532 & B1718--35 \\
J1722--3632 & B1718--36 & J1724--3149 &  & J1725--0732 &  & J1725--3546 &  & J1726--3635 &  \\
J1726--4006 &  & J1730--3350 & B1727--33 & J1731--3123 &  & J1732--3131 &  & J1732--4156 &  \\
J1734--2415 &  & J1734--3058 &  & J1735--3258 &  & J1738--2647 &  & J1738--2736 &  \\
J1739--3131 & B1736--31 & J1740--3327 &  & J1741--2054 &  & J1741--2945 &  & J1743--3153 &  \\
J1744--3922 &  & J1744--5337 &  & J1746--2849 &  & J1746--2856 &  & J1747--2809 &  \\
J1747--2958 &  & J1748--2021A & B1745--20A & J1749--2629 &  & J1749--5417 &  & J1752+2359 &  \\
J1755--0903I &  & J1755--2025 &  & J1755--2521 &  & J1755--2550 &  & J1755--26 &  \\
J1756--2225 &  & J1758--1931 &  & J1758--2540 &  & J1759--2307 &  & J1759--2549 &  \\
J1801--2154 &  & J1801--2304 & B1758--23 & J1801--2451 & B1757--24 & J1802--1745 &  & J1802--2426 &  \\
J1803--1920 &  & J1805--2032 &  & J1806--1618 &  & J1806--2125 &  & J1807+0756 &  \\
J1808--1517 &  & J1810--1441 &  & J1810--1820 &  & J1811--1736 &  & J1811--2439 &  \\
J1812+0226 & B1810+02 & J1812--1733 & B1809--176 & J1814+1130 &  & J1814--1744 &  & J1815--1738 &  \\
J1816--0755I &  & J1816--1729 & B1813--17 & J1816--5643 &  & J1818--1422 & B1815--14 & J1819--1114 &  \\
J1819--1510 &  & J1819--1717 &  & J1820--1346 & B1817--13 & J1820--1529 &  & J1821--0256 &  \\
J1821--1432 &  & J1822+1120 &  & J1822--1400 & B1820--14 & J1824--0132 &  & J1824--1118 & B1821--11 \\
J1824--1159 &  & J1824--1350 &  & J1824--1423 &  & J1825--1446 & B1822--14 & J1826--1419 &  \\
\hline
\end{tabular}
\end{center}
\end{table*}
\begin{table*}
\addtocounter{table}{-1}
\caption{Pulsars for which an RVM fit was not attempted (continued)}
\begin{center}
\begin{tabular}{cccccccccc}
\hline
\hline
JNAME & BNAME \\
\hline
J1827--0958 & B1824--10 & J1828--1007 &  & J1828--1057 &  & J1828--1101 &  & J1828--2119 &  \\
J1829+0000 &  & J1829--0734 &  & J1832--1021 & B1829--10 & J1833--0559 &  & J1834--0010 & B1831--00 \\
J1834--0426 & B1831--04 & J1834--0731 &  & J1834--0742 &  & J1835--0643 & B1832--06 & J1835--09242 &  \\
J1837--0559 &  & J1837--0604 &  & J1837--0822 &  & J1838--0453 &  & J1838--0549 &  \\
J1839--0141 &  & J1839--0223 &  & J1839--0321 &  & J1839--0332 &  & J1839--0459 &  \\
J1839--0643 &  & J1840--0559 &  & J1840--0643 &  & J1840--1419 &  & J1841--0500 &  \\
J1841--7845 &  & J1842--0153 &  & J1842--0415 &  & J1842--0800 &  & J1843+2024 &  \\
J1843--0137 &  & J1843--0355 &  & J1843--0702M &  & J1843--0744 &  & J1844--0030 &  \\
J1844--0244 & B1842--02 & J1844--0302 &  & J1844--0452 &  & J1844--0538 & B1841--05 & J1845--0826 &  \\
J1846--0257 &  & J1846--0749 &  & J1847--0443 &  & J1848+0351 &  & J1848+0647 &  \\
J1848+0826 &  & J1848--0601 &  & J1848--1243 &  & J1849+0409I &  & J1850+0423 &  \\
J1850--0006 &  & J1850--0026 &  & J1851+0233 &  & J1851--0029 &  & J1851--0114 &  \\
J1852+0008 &  & J1852+0013 &  & J1852--0127 &  & J1853+0011 &  & J1853+0505 &  \\
J1853+0545 &  & J1854+0306 &  & J1855+0205 &  & J1855+0527 &  & J1855+0700 &  \\
J1856+0102 &  & J1856+0245 &  & J1857+0143 &  & J1857+0300 &  & J1857+0526 &  \\
J1857+0809 &  & J1858+0319 &  & J1858+0346 &  & J1859+0601 &  & J1859+0603 &  \\
J1900+0438 &  & J1900+0634 &  & J1901+0459 &  & J1901+0511 &  & J1902--1036 &  \\
J1903+0601 &  & J1903+0654 &  & J1903--0258 &  & J1904+0738 &  & J1905+0600 &  \\
J1905+0902 &  & J1906+0414 &  & J1906+0509 &  & J1906+0746 &  & J1906+0912 &  \\
J1906+1854 &  & J1907+0255 &  & J1907+0602 &  & J1907+0731 &  & J1907+0859 &  \\
J1907+1247 & B1904+12 & J1908+0734 &  & J1908+0833 &  & J1908+0839 &  & J1908+0909 &  \\
J1908+0916 & B1906+09 & J1908+2351 &  & J1909+0641 &  & J1909+0912 &  & J1909+1148 &  \\
J1910+0225 &  & J1910+0517 &  & J1913+0657 &  & J1913+0832I &  & J1913+1011 &  \\
J1913+1050 &  & J1913+1145 &  & J1913+1330 &  & J1914+0631 &  & J1915+0838 &  \\
J1915+1410 &  & J1916+0844 &  & J1916+1030 & B1913+105 & J1916+1225 &  & J1917+1737 &  \\
J1918+1541I &  & J1919+1314 &  & J1919+1645 &  & J1921+1544 &  & J1922+1131 &  \\
J1926+0737 &  & J1926+1928 & B1924+19 & J1926--1314 &  & J1927+0911 &  & J1927+1856 & B1925+188 \\
J1928+1923 &  & J1929+1357 &  & J1929+1844 & B1926+18 & J1930+1852 &  & J1931+1952 &  \\
J1933+0758 &  & J1935+1745 & B1933+17 & J1941+1026 &  & J1942+1743 & B1939+17 & J1945+1834 & B1943+18 \\
J1947+1957 &  & J1948+2333 &  & J1953+1149 &  & J1954+1021 &  & J1954+2407 &  \\
J2013--0649 &  & J2027+2146 & B2025+21 & J2040+1657 &  & J2048+2255 &  & J2051+1248 &  \\
J2053+1718 &  & J2151+2315 &  & J2205+1444 &  & J2243+1518 &  &  \\
\hline
\end{tabular}
\end{center}
\end{table*}

% Don't change these lines
\bsp	% typesetting comment
\label{lastpage}
\end{document}